\documentclass[aps,11pt,tightenlines,nofootinbib,showpacs,showkeys,eqsecnum]{revtex4}
\usepackage{amsmath,amscd,amssymb}
\usepackage{color,epsfig}

\def\be{\begin{equation}}
\def\ee{\end{equation}}
\def\bea{\begin{eqnarray}}
\def\eea{\end{eqnarray}}

\newcommand{\fract}[2]{\mbox{\small $\frac{#1}{#2}$}}

\begin{document}

\title{Bound State versus Collective Coordinate Approaches in 
Chiral Soliton Models and the Width of the $\Theta^+$ Pentaquark}

\author{H. Walliser and H. Weigel}

\affiliation{Fachbereich Physik, Siegen University,  
D--57068 Siegen, Germany}

\begin{abstract}
We thoroughly compare the bound state and rigid rotator approaches 
to three--flavored chiral solitons. We establish that these two
approaches yield identical results for the baryon spectrum
and kaon--nucleon $S$--matrix in the limit that the number of 
colors ($N_C$) tends to infinity. After proper subtraction of
the background phase shift the bound state approach indeed 
exhibits a clear resonance behavior in the strangeness $S=+1$
channel. We present a first dynamical calculation of the widths
of the $\Theta^+$ and $\Theta^*$ pentaquarks for finite $N_C$ in
a chiral soliton model.
\end{abstract}


\maketitle

\section{Introduction}

Chiral soliton model predictions for the mass of the lightest
exotic pentaquark, the $\Theta^+$ with zero isospin and unit strangeness,
have been around for about two decades~\cite{Ma84,Bi84,Ch85,Pr87,Wa92,Wa92a}. 
Nevertheless, the study of such pentaquarks as potential baryon resonances 
became popular only a short while ago when experiments~\cite{Na03} indicated 
their existence\footnote{For recent reviews on the experimental situation we 
refer to refs.~\cite{Hi04,Ka05}. The numerous theory papers that have 
appeared since then may be downloaded from
{\tt http://www.rcnp.osaka-u.ac.jp/\~{}hyodo/research/Thetapub.html}.}.
These experiments were stimulated by a chiral soliton model estimate 
that claimed that the width of the pentaquark decaying into a kaon and 
a nucleon would be much smaller than those typical for hadronic decays 
of baryon resonances~\cite{Di97,Pr03,Pr04}, however, see also~\cite{Ja04}. 
Such narrow resonances could have escaped detection in earlier analyses.

\subsection{Hadronic Decays in Soliton Models}

Estimates for the $\Theta^+$ width are based on identifying the 
interpolating kaon field with the (divergence) of the axial current, which 
in turn is computed solely from the classical soliton. This is essentially 
a generalization to flavor $SU(3)$ of the early calculation of the 
$\Delta\to \pi N$ decay width in the pioneering paper of ref.~\cite{Ad83} in
the two flavor Skyrme model. 
Since then there have been quite some efforts to understand that process 
in chiral soliton models by going beyond the simple identification of 
the pseudoscalar meson fields with the axial 
current~\cite{Ue85,Sa87,Ho87,Ve88,Di88,Ho90,Ho90a,Ha90,Do94};
see also chapter IV in ref.~\cite{Ho93}. It must be stressed that 
the description of hadronic decays in soliton models is a severe problem
and has become known as the \emph{Yukawa} problem in soliton models.
The reason is that the interaction Hamiltonian that mediates such hadronic 
transitions must be linear in the field operator that corresponds to the 
pseudoscalar meson in the final state. On the other hand the na{\"\i}ve 
definition of the soliton as a stationary point of the action prohibits 
such a linear term. Hence in these models resonance widths must be 
extracted from appropriate $S$--matrix elements of meson baryon scattering.
Most of the studies on meson baryon scattering are restricted to the
adiabatic approximation~\cite{Wa84,Ha84,Kr86,Schw89} which represents
the next--to--leading order in the $1/N_C$ expansion where $N_C$ is the 
number of colors. Unfortunately in the large--$N_C$ limit the $\Delta$
is degenerate with the nucleon and its width is of even lower 
order in that expansion. Therefore its width cannot be computed in
adiabatic approximation and the above cited studies mainly motivate 
Born contributions to the $S$--matrix element of pion nucleon scattering 
in the $P_{33}$ channel at ${\cal O}(1/N_C)$. For the pentaquarks
the situation is very different. As we will see, the $\Theta^+$--nucleon
mass difference is non--zero even as $N_C\to\infty$ and hence its
hadronic decay width should be at the same order as the adiabatic
approximation for kaon--nucleon scattering. In that respect the
hadronic decays of pentaquarks are a perfect playground as 
any treatment that induces Born contributions can be tested against the
adiabatic approximation. 

\subsection{Quantizing the Soliton in Flavor $\mathbf{SU(3)}$}

When treating chiral solitons in flavor $SU(3)$ we face the problem 
how to properly generate baryon states from the classical 
soliton. Essentially there are two methods at our disposal,
the bound state approach (BSA)~\cite{Ca85} and the rigid rotator 
approach (RRA)~\cite{Gu84}; see {\it e.g.\@} ref.~\cite{We96} for 
a review and further references. The latter is essentially the
generalization of the collective coordinate quantization adopted 
earlier for the two flavor model~\cite{Ad83}. In this approach collective
coordinates that describe the spin flavor orientation of the soliton
are introduced and quantized canonically. As this corresponds 
to quantizing a rigid top in $SU(3)$ it is natural that the 
corresponding eigenstates are characterized according to $SU(3)$
multiplets, most prominently the octet ($\mathbf{8}$) and 
decuplet ($\mathbf{10}$) for baryons with spin $J=\frac{1}{2}$ and
$J=\frac{3}{2}$, respectively. However, also more exotic $SU(3)$
representations like the anti--decuplet ($\overline{\mathbf{10}}$) 
emerge. In addition to states with quantum numbers of octet baryons this
representation also contains states with quantum numbers that 
\emph{cannot} be built as three--quark composites. These states 
contain (at least) one additional quark--antiquark pair which lead
to the notion of exotic pentaquarks. Most prominently there have 
been experimental claims~\cite{Na03,Hi04,Ka05} for a $\Theta^+$
pentaquark, the strangeness~$S=+1$ isosinglet that sits at the top 
of the anti--decuplet.
 
In the BSA kaon fluctuations $\eta_\alpha(\vec{x\,},t)$, with 
$\alpha=4,\ldots,7$, are introduced and quantized as harmonic 
oscillations. It corresponds to the adiabatic
approximation and should be exact as $N_C\to\infty$. Common to all 
soliton models is the appearance of a bound state with strangeness~$S=-1$ 
which serves to construct states that correspond to the ordinary hyperons, 
$\Lambda$, $\Sigma$ and $\Sigma^*$. On the other hand neither a bound 
state nor a clear resonance has been observed in the 
$S=+1$ channel. This non--observation has often been used to argue that 
the pentaquark prediction would be a mere artifact of the 
RRA~\cite{It03,Co03,Ch05}. Here we will show that this statement 
is completely based on a misinterpretation of the RRA. For a sensible 
and ultimate comparison of the BSA and the RRA it is necessary to also 
include small amplitude fluctuations in the RRA. We call that approach
the rotation--vibration approach (RVA). However, the fluctuations in the 
RVA must be constrained to the subspace that is orthogonal to the rigid 
rotation. We therefore label these fluctuation by 
$\overline{\eta}_\alpha(\vec{x\,},t)$. Then the test mentioned at the 
end of subsection A is to first compute the $S$--matrix from the 
constrained fluctuations, $\overline{\eta}_\alpha(\vec{x\,},t)$ and 
supplement that result with the contributions from the (iterated)
$\Theta^+$ exchange. That yields the wave--functions 
$\widetilde{\eta}_\alpha(\vec{x\,},t)$ that are constrained to the 
same subspace as $\overline{\eta}_\alpha(\vec{x\,},t)$. The
$\widetilde{\eta}_\alpha(\vec{x\,},t)$ are the full solution to the RVA.
In the limit $N_C\to\infty$~ $\widetilde{\eta}_\alpha(\vec{x\,},t)$
must yield the same $S$--matrix as in the BSA, {\it i.e.\@} 
when extracted from ${\eta}_\alpha(\vec{x\,},t)$. In performing this 
test, our studies serve three major purposes: First, we will
establish the equivalence of BSA and RVA in the large--$N_C$ limit.
Second, we will compare $\widetilde{\eta}_\alpha$ and $\overline{\eta}_\alpha$
to extract a distinct $S=1$ resonance in kaon nucleon scattering that 
survives as $N_C\to\infty$. Third, we will provide the \emph{first} 
dynamical calculation of the $\Theta^+$ width in a chiral soliton model
for finite $N_C$.

\subsection{General Remarks}

Let us round off this introduction section with a few general 
remarks on the comparison between the BSA, the RRA and the RVA.
Generally we may always introduce collective coordinates to 
investigate specific excitations. This is irrespective of 
whether the corresponding excitation energies are suppressed 
in the large--$N_C$ limit or not. That is, in contrast to the 
claims of refs.~\cite{Co03} no separation of 
scales is required to validate the collective
coordinate quantization. Eventually the coupling terms
and constraints ensure correct results. Even more, we will 
observe in section~V that collective coordinates may also
be introduced for $m_K\ne m_\pi$. This result generalizes
to the statement that collective coordinates are \emph{not}
necessarily linked to the zero--modes of a system and any
distinction between dynamical and static zero--modes~\cite{Ch05} 
is obsolete. Already some time ago the concept of 
describing baryon resonances as collective excitations 
has been successfully applied to quadrupole~\cite{Ha84a}
and monopole excitations, both in two~\cite{Ha84b} and
three~\cite{Sch91a} flavor models. In particular collective
monopole excitations have been studied in the context of
the pentaquark excitation~\cite{We98,We04}.

Sometimes the correct description of the $\Lambda(1405)$ as an 
$S$--wave bound state is considered as an argument in favor 
of the BSA vs. the RRA. Again it must be stressed that the full 
RVA also contains kaon fluctuations and thus also describes the 
formation of an $S$--wave bound state. As shown in ref.~\cite{Schw91} 
this properly accounts for the $\Lambda(1405)$ in the RVA. 

\subsection{Outline}

This paper is organized as follows. In section II we
will define the model and describe the field parameterization in
the bound state and rigid rotator approaches in more detail.
In sections III and IV we extensively compare these approaches and show
how they yield identical spectra in the limit $N_C\to\infty$,
even for non--zero symmetry breaking, when the fluctuating modes are
limited to the same subspace. In section III we also show that 
the constraints for $\overline{\eta}_\alpha$ serve to separate 
resonance and background pieces of the phase shift. Section V 
represents a main piece of our paper as we combine rotational and 
fluctuation degrees of freedom to construct the solution for the 
RVA up to quadratic order in the fluctuations. We employ this solution 
to compute the width of the $\Theta^+$ pentaquark for $m_K=m_\pi$ 
in section VI. Another major result is that in the flavor symmetric 
case only a \emph{single} collective coordinate operator mediates 
the transition $\Theta^+\to KN$ at the leading order of the large--$N_C$
expansion. This is in direct contradiction to approaches that construct 
transition interactions with (at least) two $SU(3)$ operators and adjust 
the coefficients to yield a narrow $\Theta^+$~\cite{Di97,Pr03,El04,Pr05}. 
In section VII we extend our treatment to the physical case 
$m_K\ne m_\pi$. This brings a second $SU(3)$ operator into the game. In
section VIII we collect our numerical results and summarize in section IX. 
We relegate technical details on the derivation of the equation of motion 
for the constrained fluctuations to an appendix.

\section{The Model}
 
For simplicity we consider the Skyrme model~\cite{Sk61} as a 
particular example for chiral soliton models. However, we stress that our 
qualitative results do indeed generalize to \emph{all} chiral soliton models 
because these results solely originate from the treatment of the model 
degrees of freedom. 

A preliminary remark on notation: In what follows we adopt the convention 
that repeated indices are summed over in the range 
\begin{equation}
\begin{array}[b]{lcll}
a,b,c,\ldots &=& 1,\ldots,8 & \cr
\alpha,\beta,\gamma,\ldots &=& 4,\ldots,7 & {\rm (kaonic)}\cr
i,j,k,\ldots &=& 1,2,3 \hspace{2cm} &  {\rm (pionic)}\,.
\end{array}
\label{convention}
\end{equation}

Chiral soliton models are functionals of the chiral field, $U$,
the non--linear realization of the pseudoscalar mesons, $\phi_a$
\be
U(\vec{x\,},t)={\rm exp}\left[\frac{i}{f_\pi}
\phi_a(\vec{x\,},t)\lambda_a\right]\,,
\label{chiralfield}
\ee
with $\lambda_a$ being the Gell--Mann matrices of $SU(3)$. For a convenient 
presentation of the model we split the action into three pieces
\be
\Gamma = \Gamma_{SK} + \Gamma_{WZ} + \Gamma_{SB} \,.
\label{lag}
\ee
The first term represents the Skyrme model action
\be
\Gamma_{SK} =
\int d^4 x\, {\rm tr}\,\left\{ \frac{f^2_\pi}{4} 
\left[ \partial_\mu U \partial^\mu U^\dagger \right]
+ \frac{1}{32\epsilon^2} \left[ [U^\dagger \partial_\mu U , 
U^\dagger \partial_\nu U]^2\right] \right\} \, .
\label{Skmodel}
\ee
Here $f_\pi=93{\rm MeV}$ is the pion decay constant and $\epsilon$ is 
the dimensionless Skyrme parameter. In principle this is a free 
model parameter. The two--flavor version of the Skyrme model suggests to 
put $\epsilon=4.25$ from reproducing the $\Delta$--nucleon mass 
difference. The anomaly structure of QCD is incorporated via the 
Wess--Zumino action~\cite{Wi83} 
\be
\Gamma_{WZ} = - \frac{i N_C}{240 \pi^2}
\int d^5x \ \epsilon^{\mu\nu\rho\sigma\tau}
{\rm tr}\,\left[ L_\mu L_\nu L_\rho L_\sigma L_\tau\right] \,,
\label{WZ}
\ee
with $L_\mu = U^\dagger \partial_\mu U$. Note that 
$\Gamma_{WZ}$ vanishes in the two--flavor version of the model.
Furthermore $N_C$ is the number of colors, the hidden expansion
parameter of QCD~\cite{tH74,Wi79}. 

The flavor symmetry 
breaking terms are contained in $\Gamma_{\rm SB}$ 
\be
\Gamma_{SB} = \frac{f_\pi^2}{4}\, \int d^4x\,
{\rm tr}\, \left[{\cal M}\left(U + U^\dagger - 2 \right)\right]
\qquad {\rm with} \qquad
{\cal M}=
\begin{pmatrix}
m_\pi^2 & 0 & 0\cr
0 &  m_\pi^2 & 0 \cr
0 & 0& 2m_K^2-m_\pi^2
\end{pmatrix}\,.
\label{SB}
\ee
Unless otherwise stated we will adopt the empirical data for the masses of 
the pseudoscalar mesons: $m_\pi=138{\rm MeV}$ and $m_K=495{\rm MeV}$. 
We do not include terms that distinguish between pion and
kaon decay constants even though they differ by about 20\% empirically.
This is a mere matter of convenience rather than negligence. This
omission leads to a considerable underestimation of symmetry breaking 
effects~\cite{We90} which approximately can be accounted for by rescaling 
the kaon mass $m_K\to m_Kf_K/f_\pi$.

Although physically sensible
results are obtained by setting $N_C=3$, it is very
illuminating to organize the calculation in powers of $1/N_C$
and consider $N_C$ undetermined.
In the context of this $N_C$ counting the mass parameters 
are ${\cal O}(N_C^0)$ while $f_\pi={\cal O}(\sqrt{N_C})$
and $\epsilon={\cal O}(1/\sqrt{N_C})$ to ensure that 
the (perturbative) $n$--point functions scale as  
$N_C^{1-n/2}$~\cite{Wi79}. In order to study the
$N_C$ dependence in the three flavor version of the soliton 
model we include the $N_C$ dependence in the choice of 
the model parameters via
\be
f_\pi=93{\rm MeV}\sqrt{\frac{N_C}{3}}\quad {\rm and} \quad
\epsilon=4.25\sqrt{\frac{3}{N_C}}\,.
\label{parameters}
\ee

The action, eq.~(\ref{lag}) allows for a topologically non--trivial
classical solution, the famous hedgehog soliton
\be
U_0(\vec{x\,})={\rm exp}\left[i\vec{\lambda\,}\cdot\hat{x} F(r)\right]\,,
\quad r=|\vec{x\,}|
\label{hedgehog}
\ee
which is embedded in the isospin subspace of flavor $SU(3)$
parameterized by the Gell--Mann matrices $\lambda_i$, $i=1,2,3$. Other 
embeddings are possible but energetically disfavored due to flavor 
symmetry breaking. The chiral angle, $F(r)$ is obtained from integrating 
the classical equation of motion extracted from eqs.~(\ref{Skmodel}) 
and~(\ref{SB}) subject to the boundary
condition\footnote{The overall sign is a matter of convention.}
$F(0)-F(\infty)=\pi$ ensuring unit winding (baryon) number. The soliton 
can be constructed as a function of the dimensionless variable 
$\epsilon f_\pi r$ and is thus not subject to $N_C$ scaling.

The excitations of this soliton configuration are quantized 
to generate physical baryon states. The central task in 
extending this picture to three flavors is the incorporation of 
strange degrees of freedom, that is {\it kaons}. In a first step we 
introduce fluctuations\footnote{In what follows we will discuss 
different parameterizations of the chiral field, $U$. We abstain from 
introducing additional labels to distinguish among them.}
$\eta_a(\vec{x\,},t)$ with $a=1,\ldots,8$, for the pseudoscalar
fields~\cite{Ca85,Ca88,Kl89},
\be
U(\vec{x\,},t)=  \sqrt{U_0(\vec{x\,})}\,
{\rm exp}\left[\frac{i}{f_\pi}\lambda_a\eta_a(\vec{x\,},t)\right]
\sqrt{U_0(\vec{x\,})}\,.
\label{ckfluct}
\ee
Expanding the action in powers of these fluctuations actually is 
an expansion in $\eta_a/f_\pi$ and thus a systematic series 
in $1/\sqrt{N_C}$. The term quadratic in $\eta_a$ is of special 
interest. It describes scattering of mesons off a potential that 
is generated by the classical soliton, eq.~(\ref{hedgehog}). In 
particular an anti--kaon is bound in the $P$--wave channel. Combined 
with the soliton it may be interpreted as the $\Lambda$ hyperon. 
However, the system consisting of the soliton and this bound anti--kaon 
has still to be projected onto states with good spin and isospin.
This is accomplished by introducing collective coordinates 
for the spin--isospin orientation\footnote{Due to the hedgehog 
symmetry only one set of collective coordinates is required.}, 
$A_2(t)\in SU(2)$
\be
U(\vec{x\,},t)=A_2(t) \sqrt{U_0(\vec{x\,})}\,
{\rm exp}\left[\frac{i}{f_\pi}\lambda_a \eta_a(\vec{x\,},t)\right]
\sqrt{U_0(\vec{x\,})}\, A_2^\dagger(t)\,
\label{Usu2}
\ee
and quantizing them canonically~\cite{Ca85,Kl89}. In this treatment
the kaon fluctuations $a=\alpha=4,\ldots,7$ decouple in the
adiabatic equations of motion. This leads to the so--called bound 
state approach (BSA) that we will review in section III. 

The seemingly alternative approach is to allow the collective coordinates 
to span the \emph{full} flavor space, $A_3(t)\in SU(3)$ and quantize $A_3$ 
canonically to generate baryon states with non--zero strangeness
while omitting the fluctuations. 
Baryon states are then classified according to $SU(3)$ representations.
This is the so--called rigid rotator approach (RRA)~\cite{Gu84}. 
However, for a sensible investigation of the rotation--vibration
coupling the collective coordinate approach must be completed to 
also contain meson fluctuations. Hence the {\it ansatz} that defines 
the RVA reads
\be
U(\vec{x\,},t)=A_3(t) \sqrt{U_0(\vec{x\,})}\,
{\rm exp}\left[\frac{i}{f_\pi}\lambda_a\widetilde{\eta}_a(\vec{x\,},t)\right]
\sqrt{U_0(\vec{x\,})}\, A_3(t)^\dagger\,.
\label{Usu3}
\ee
Obviously eq.~(\ref{Usu3}) comprises the {\it ansatz}, eq.~(\ref{Usu2}).
In contrast to eq.~(\ref{Usu2}) the {\it ansatz} (\ref{Usu3}) has 
additional collective rotations into strange directions. But these field 
components are also contained in the fluctuations, $\widetilde{\eta}_\alpha$.
It is therefore necessary to constrain these fluctuations to the 
subspace that is orthogonal to the rigidly rotating hedgehog.
These constraints modify the equations of motion for the fluctuations,
such that $\widetilde{\eta}_\alpha\ne\eta_\alpha$. This is not new:
In principle we encounter the same problem already in $SU(2)$ where
the {\it ans\"atze} eqs.~(\ref{Usu2}) and~~(\ref{Usu3}) are identical.
The difference to the $SU(3)$ case is that the redundant modes are 
automatically orthogonal to the rigid rotations when treating the 
$SU(2)$ model in adiabatic approximation. This changes, however, as 
soon as non--adiabatic contributions or external fields {\it e.g.\@} 
photons~\cite{Me97}, are considered. 
We will discuss and derive the modifications that occur in the $SU(3)$ 
case in section III and a separate appendix, respectively.

It is obvious that the 
{\it ans\"atze}~(\ref{Usu2}) and~(\ref{Usu3}) span the same Fock space 
in the baryon number one sector. An essential difference,
however, is that an expansion to a given order in $\eta_a$ 
using eq.~(\ref{Usu2}) contains the kaon fields only up to a fixed
power in the $1/N_C$ expansion while the {\it ansatz}~(\ref{Usu3})
treats the collective rotations to \emph{all} orders. Obviously the 
two treatments must yield identical results in the limit 
$N_C\to\infty$. This provides a thorough check on our calculations. 
We also stress that the RVA formulation induces terms in the action 
that are linear in $\widetilde{\eta}_\alpha$
(in contrast the BSA does not give terms linear in $\eta_\alpha$).
This gives rise to Yukawa couplings from which we will evaluate 
widths of rotational excitations.

There have been earlier studies of the non--adiabatic 
rotation--vibration coupling in the Skyrme model. In the
$SU(2)$ version the contribution linear in the time derivative 
of $A_2$ has been shown to properly describe the Tomozawa--Weinberg
relations for $S$--wave pion nucleon scattering~\cite{Wa91}.
Also in the two flavor model, the terms quadratic in the 
time derivative of the collective coordinates have been employed
to incorporate thresholds and $\Delta$ resonance exchange effects
in pion--nucleon scattering~\cite{Ha90,Ho90,Ho90a,Schw89}.
In the three--flavor BSA the rotation--vibration coupling
gives rise to the hyperfine mass splitting~\cite{Kl89} that we will 
discuss in the following section. Also in $SU(3)$ meson baryon 
scattering has been investigated in the RVA~\cite{Schw91}. Unfortunately,
the strangeness, $S=+1$ channel, which contains the $\Theta^+$
pentaquark, was not considered in that outstanding paper. 

\section{Bound state approach (BSA)}

In this section we will discuss the construction of $P$--wave baryons 
with strangeness $S=\pm1$ in the framework of the BSA starting
from the parameterization, eq.~(\ref{ckfluct}). Baryons with 
zero strangeness are treated as in two--flavor chiral soliton 
models~\cite{Ad83}. 

\subsection{Conventional BSA in the P--wave channel}

As in refs.~\cite{Ca85,Ca88} we expand the action to quadratic order in 
the kaonic fluctuations $\eta_\alpha$ in the background of the 
soliton, eq.~(\ref{hedgehog}). These wave--functions describe 
kaon--nucleon scattering in the intrinsic frame. In this frame the modes
decouple with respect to grand spin ${\cal G}=\ell\pm\fract{1}{2}$, 
the eigenvalue of the sum of kaon orbital angular momentum 
($\ell$) and isospin.  This defines the 
intrinsic $T$--matrix, $T_{{\cal G},\ell}$. The $T$--matrix elements 
in the laboratory frame $T(\ell I2J)$ are linear combinations of the
$T_{{\cal G},\ell}$~\cite{Ha84,Ma88}. For the $\Theta^+$ channel
($I=0$, $J=\frac{1}{2}$) this linear relation is very simple
\be
T(P01) = T_{\frac{1}{2}1}
\label{pwaveint}
\ee
{\it i.e.\@} only the intrinsic $P$--wave kaon with 
grand spin ${\cal G}=\frac{1}{2}$ contributes.
Its wave--function is characterized by a single 
radial function, that parametrically depends on the frequency in a Fourier
analysis
\be
\begin{pmatrix}
\eta_4+i\eta_5\cr
\eta_6+i\eta_7
\end{pmatrix}_{P}(\vec{x\,},t)\,
=\int_{-\infty}^\infty\, d\omega\, {\rm e}^{i\omega t}\,
\eta_\omega(r)\,\hat{x}\cdot\vec{\tau}\,\chi(\omega)\,.
\label{pwave}
\ee
Here $\chi(\omega)$ is a two--component iso--spinor that
contains the amplitudes of the kaon mode. Upon quantization its
components will eventually be elevated to creation-- and annihilation 
operators.  Of course, an analogous decomposition exists for 
$\widetilde{\eta}_\alpha$ that defines the radial function 
$\widetilde{\eta}_\omega(r)$. In what follows we will omit the 
subscript $\omega$. The amplitude $\chi$ drops out from the 
equation of motion that can be formulated for the radial functions,
\bea
\Pi(r)&=&\left[\omega M_K(r)-\lambda(r)\right]\eta(r)\,,\cr\cr
h^2\eta(r)+\frac{\lambda^2(r)}{M_K(r)}&=&
\left[\omega -\frac{\lambda(r)}{M_K(r)}\right]\Pi(r)\,.
\label{blaizoteqm}
\eea
Up to a factor $i$ the radial function $\Pi(r)$ parameterizes the
momentum conjugate to the $P$--wave fluctuation~\cite{Bl88}.
The radial functions in eq.~(\ref{blaizoteqm}) can be taken from the 
literature~\cite{Ca85,Ca88,Scha94} in terms of the chiral angle $F=F(r)$,
\bea
\lambda(r)&=&\frac{N_C}{4f_\pi^2}B_0(r)
=-\frac{N_C}{8\pi^2f_\pi^2}F^\prime\,\frac{{\rm sin}^2F}{r^2}\,,\quad
M_K(r)=1+\frac{1}{4f_\pi^2\epsilon^2}
\left[F^{\prime2}+2\frac{{\rm sin}^2F}{r^2}\right]\,\cr\cr
h^2\,&=&h^2_0\,+m_K^2-m_\pi^2\quad {\rm with} \quad
h^2_0\,=-\frac{d^2}{dr^2}-\frac{2}{r}\frac{d}{dr}+V_{\rm eff}(r)\,,
\label{radfct}
\eea
where $F^\prime=\frac{dF(r)}{dr}$. The effective potential $V_{\rm eff}(r)$
is quite an involved function that also contains spatial derivative 
operators. Since it does not play any special role
for our studies we do not repeat it here. The radial function $\lambda(r)$,
proportional to the baryon density, $B_0(r)$,
originates from the WZ--term, eq.~(\ref{WZ}) and thus contains the 
explicit factor $N_C$. The radial function $M_K(r)$ is the metric function
for kaon fluctuations. This becomes obvious when eliminating the momentum 
$\Pi(r)$ in favor of a second order differential equation for $\eta(r)$,
\be
h^2\,\eta(r)+\omega\left[2\lambda(r)-\omega M_K(r)\right]\eta(r)=0\,,
\label{scndorder}
\ee
as $M_K(r)$ multiplies the term of highest power in the frequency, $\omega$.
Eq.~(\ref{scndorder}) could have been easily obtained without reference
to the conjugate momentum. However, we have introduced it here because
it becomes essential when implementing the constraints that
characterize $\overline{\eta}$.
The appearance of factors $N_C/f_\pi^2$ and 
$f_\pi^2\epsilon^2$ in the radial functions shows that 
$\eta(r)$ does not scale with $N_C$. Hence terms in the action that
are of higher than quadratic order in $\eta_\alpha$ are suppressed
due to the additional factor $1/f_\pi$. Thus solving eq.~(\ref{scndorder})
provides the \emph{exact} $T$--matrix in the limit $N_C\to\infty$.

The equation of motion~(\ref{scndorder}) is not invariant under particle 
conjugation $\omega\leftrightarrow-\omega$, yielding different results for 
kaons ($\omega>0$) and anti--kaons ($\omega<0$). This difference obviously
originates from the Wess--Zumino term. It is well known that 
equation~(\ref{scndorder}) has a bound state solution at 
$\omega=-\omega_\Lambda$,
that gives the mass difference between the $\Lambda$--hyperon and 
the nucleon in the large--$N_C$ limit. As this energy eigenvalue 
is negative it corresponds to a kaon, {\it i.e.\@} it carries 
strangeness $S=-1$. In the symmetric case ($m_K=m_\pi$) the bound state 
energy vanishes, {\it i.e.\@} the bound state is nothing but the zero 
mode of $SU(3)$ flavor symmetry. The radial function of the properly
normalized zero mode reads
\be
z(r)=\sqrt{4\pi}\frac{f_\pi}{\sqrt{\Theta_K}}\, {\rm sin}\frac{F(r)}{2}
\quad {\rm with} \quad
\Theta_K=f_\pi^2\int d^3r\, M_K(r)\, {\rm sin}^2\frac{F(r)}{2} 
= 1.894{\rm GeV}^{-1}\left(\frac{N_C}{3}\right)\,.
\label{zeromode}
\ee
It satisfies the important relation $h^2_0\, z(r)=0$. 
The normalization constant, $\Theta_K$ is the moment of inertia
for flavor rotations into strangeness direction.
In what follows we will employ the expressions "zero--mode"
and "rotational mode" synonymously.

When symmetry breaking is switched on, the bound state evolves 
continuously from this zero mode. For $m_K=495{\rm MeV}$ we find 
$\omega_\Lambda=196{\rm MeV}$. This bound state energy can be
reliably estimated by sandwiching the equation of motion~(\ref{scndorder})
between zero mode wave--functions and solving for $\omega$.
We first define
\be
\omega_0=\int r^2dr z^2(r)2\lambda(r)=
\frac{N_C}{4\Theta_K}=0.396{\rm GeV}
\label{defom0}
\ee
and
\be
\Gamma=\frac{8\Theta_K}{3}\left(m_K^2-m_\pi^2\right)
\int r^2 dr\, z^2(r)
=2.969{\rm GeV}^{-1}\,\left(m_K^2-m_\pi^2\right)\,\frac{N_C}{3}\,,
\label{defGamma}
\ee
which are ${\cal O}(N_C^0)$ and ${\cal O}(N_C)$, 
respectively. Then the equation for $\omega$ reads
\be
\omega^2=\frac{3\Gamma}{8\Theta_K}+\omega_0\,\omega\,.
\label{omscnd}
\ee
This quadratic equation for $\omega$ has the negative 
valued solution $-\omega_\Lambda$ with
\be
\omega_\Lambda=\frac{1}{2}\left[\sqrt{\omega_0^2+\frac{3\Gamma}{2\Theta_K}}
-\omega_0\right]\,.
\label{oml}
\ee
For $m_K=m_\pi$
we have $\Gamma=0$ and thus also $\omega_\Lambda=0$ from this estimate.
In the realistic case we find $\Gamma=671{\rm MeV}\left(\frac{N_C}{3}\right)$
yielding $\omega_\Lambda\approx217{\rm MeV}$ which is only about 
10\% larger than the energy eigenvalue of the exact solution to
eq.~(\ref{scndorder}). Obviously the energy eigenvalue $\omega_\Lambda$
is a non--linear function of flavor symmetry breaking.
In particular the observed relation 
$3\Gamma/2\Theta_K>\omega_0^2$ indicates that the perturbative 
treatment of flavor symmetry breaking is bound to fail\footnote{The 
realistic case is even worse, as $\Gamma$ is underestimated by about 50\%
in our approximation with $f_K=f_\pi$.}.

\begin{figure}
\centerline{
\epsfig{file=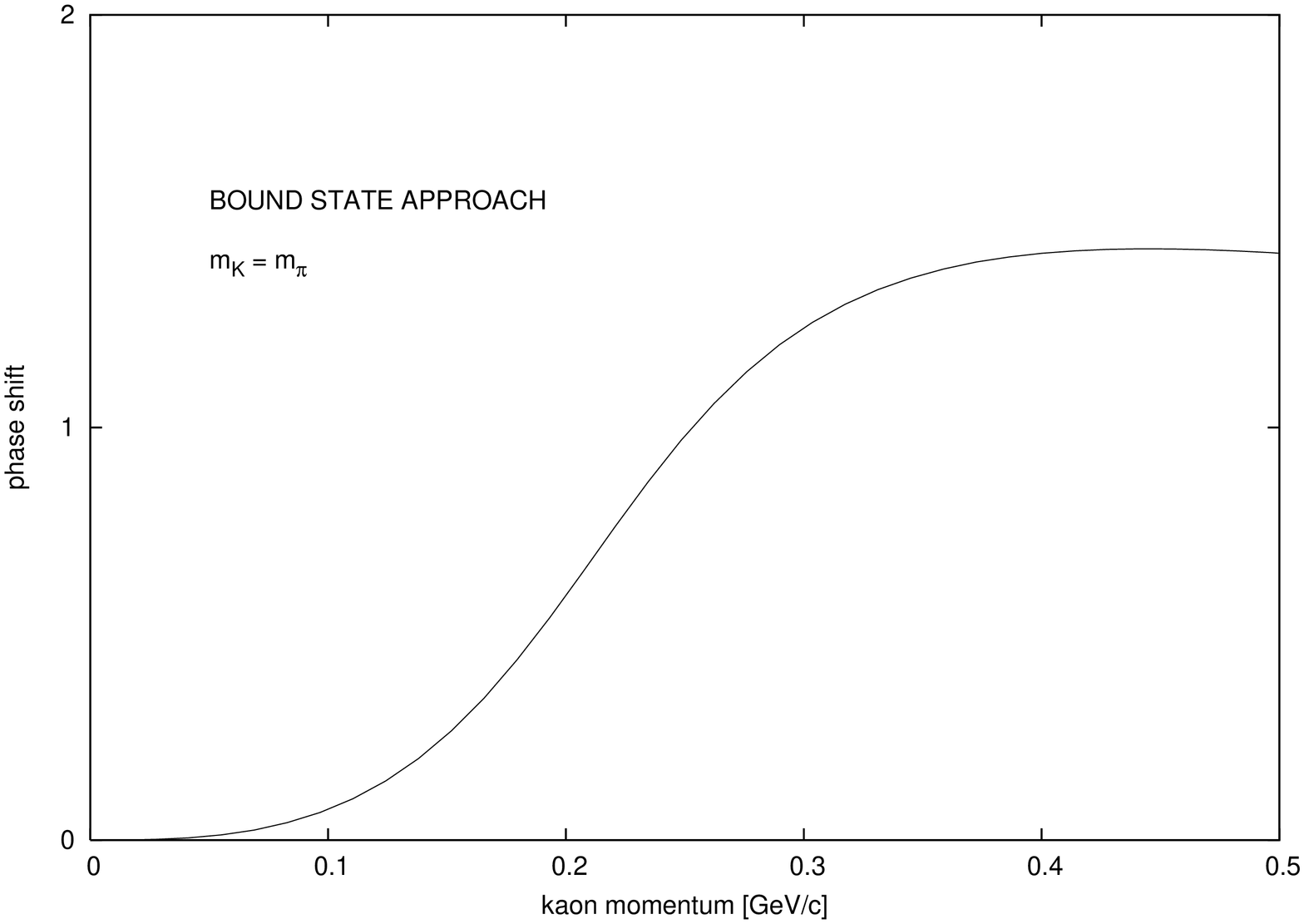,width=5cm,height=6cm,angle=270}\hspace{2cm}
\epsfig{file=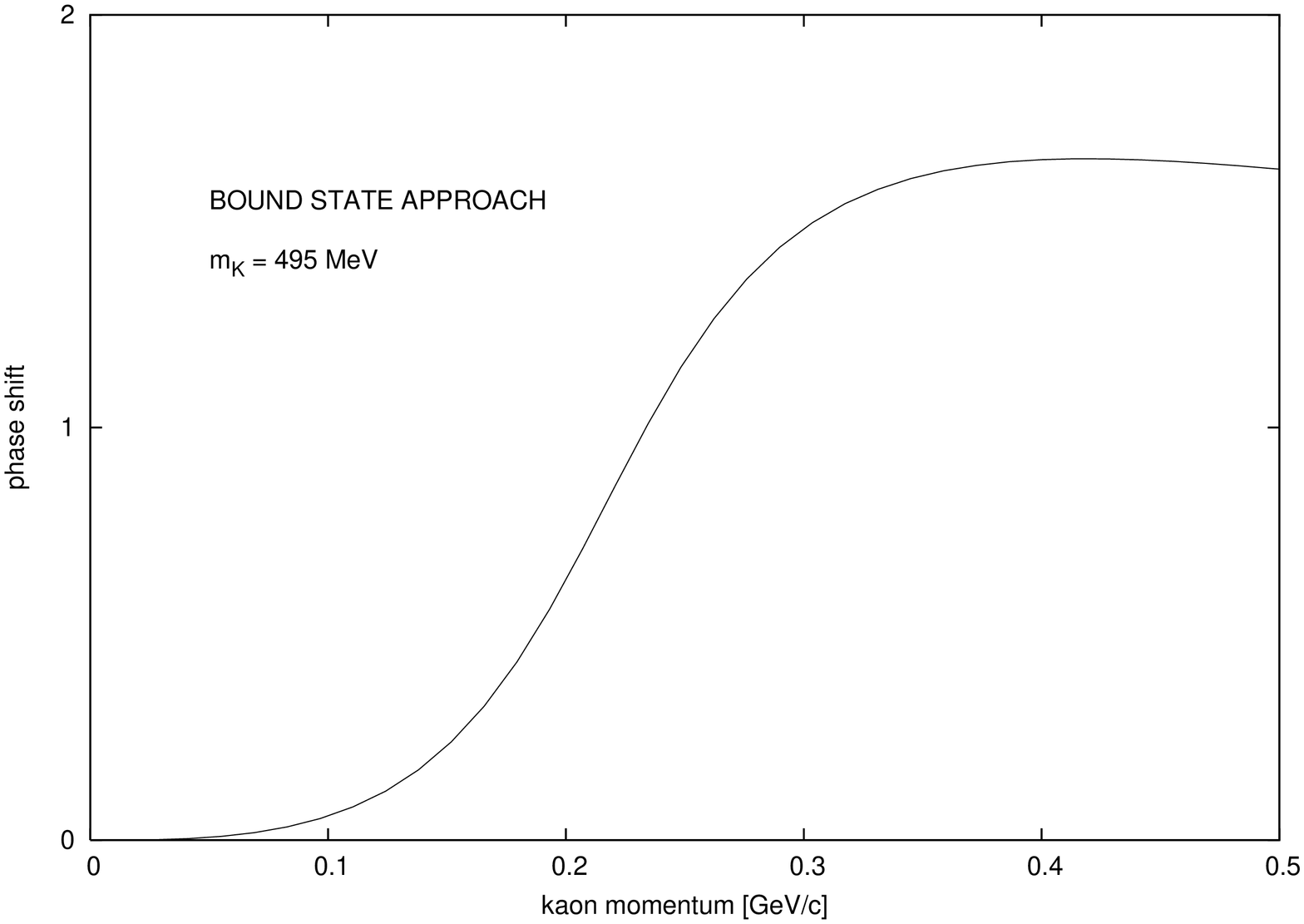,width=5cm,height=6cm,angle=270}}
\caption{\label{fig_1}The phase shift in the strangeness $S=+1$
channel for $m_K=m_\pi$ (left panel) and $m_K\ne m_\pi$ (right panel)
as functions of the kaon momentum $k^2=\omega^2-m_K^2$ in the BSA.}
\end{figure}

The quadratic equation~(\ref{omscnd}) has a second solution: $\omega_\Theta$.
It corresponds to strangeness $S=+1$ fluctuations and determines the 
mass difference between the $\Theta^+$ pentaquark and the nucleon,
\be
\omega_\Theta=\frac{1}{2}\left[\sqrt{\omega_0^2+\frac{3\Gamma}{2\Theta_K}}
+\omega_0\right]\,.
\label{omt}
\ee
Also this eigenmode evolves from the zero mode. However, the Wess--Zumino 
terms shifts  $\omega_\Theta$ to the positive continuum. We therefore
expect a resonance structure in the phase shift around 
$\omega_\Theta=396{\rm MeV}$ and $617{\rm MeV}$ for the 
cases $m_K=m_\pi$ and $m_K\ne m_\pi$,
respectively. These phase shifts are shown in figure~\ref{fig_1}.
Apparently no clear resonance structure is visible; the phase shifts 
hardly reach $\pi/2$ at the estimated energies. The absence of such a 
resonance has previously lead to the conjecture and criticism that 
there would not exist a bound pentaquark in the
large--$N_C$ limit~\cite{It03} and that it would be a mere
artifact of the RRA. The absence of pentaquarks in the BSA does
not seem to be a result limited to the simple Skyrme model.
For example, in the BSA to the NJL chiral quark soliton model
a pentaquark bound state has not been observed either~\cite{We93}.

We have obtained the above equations for $\omega_\Lambda$ and
$\omega_\Theta$ by projecting the BSA equation~(\ref{scndorder})
onto the rotational zero mode. Hence they are exact when the
fluctuations are constrained to the rotational subspace.

\subsection{Fluctuations orthogonal to rigid rotations}

Similar to the above study we eventually have to consider small 
amplitude fluctuations about the rotating soliton in the RVA, {\it i.e.\@}
$\widetilde{\eta}_\alpha$ in eq.~(\ref{Usu3}) in sections V--VIII. In 
these sections we will show that the phase shifts extracted from 
$\widetilde{\eta}$ 
and $\eta$ are identical in the large--$N_C$ limit; the ultimate test on 
the RVA. In contrast to the BSA fluctuations $\eta_\alpha$ explored so far, 
the $\widetilde{\eta}_\alpha$ must be constrained to the subspace orthogonal 
to the rigid rotations of the hedgehog to avoid double--counting. These 
constraints follow directly from the {\it ansatz}, eq.~(\ref{Usu3}) and 
we will derive them in the appendix. Here we will study the solutions 
to the equations of motion that arise  when the BSA equations of 
motion~(\ref{blaizoteqm}) are supplemented by these constraints. We call 
these solutions $\overline{\eta}$ and $\overline{\Pi}$. 

As noted above, the RVA Lagrangian contains terms that are linear in 
$\widetilde{\eta}$ providing Yukawa couplings to rotational excitations 
of the soliton. The equation of motion for $\widetilde{\eta}$ will then 
differ from the one for $\eta$ in two aspects: (i)~the constraints and 
(ii)~exchanges of resonances induced by the Yukawa couplings. 
The important point here is obvious: supplementing the equation of motion 
for $\eta$ by the constraints or omitting the Yukawa exchange terms 
from the one for $\widetilde{\eta}$ identically gives the one for the 
auxiliary field $\overline{\eta}$. Since $\overline{\eta}$ is obtained 
by removing the resonance pieces from $\widetilde{\eta}$ it is the
background contribution to the scattering amplitude. This strongly
motivates the study of the auxiliary field $\overline{\eta}$ as an 
intermediate step to construct the RVA fluctuations, $\widetilde{\eta}$;
even though the introduction of $\overline{\eta}$ seems somewhat 
artificial at this moment.

\begin{figure}
\centerline{
\epsfig{file=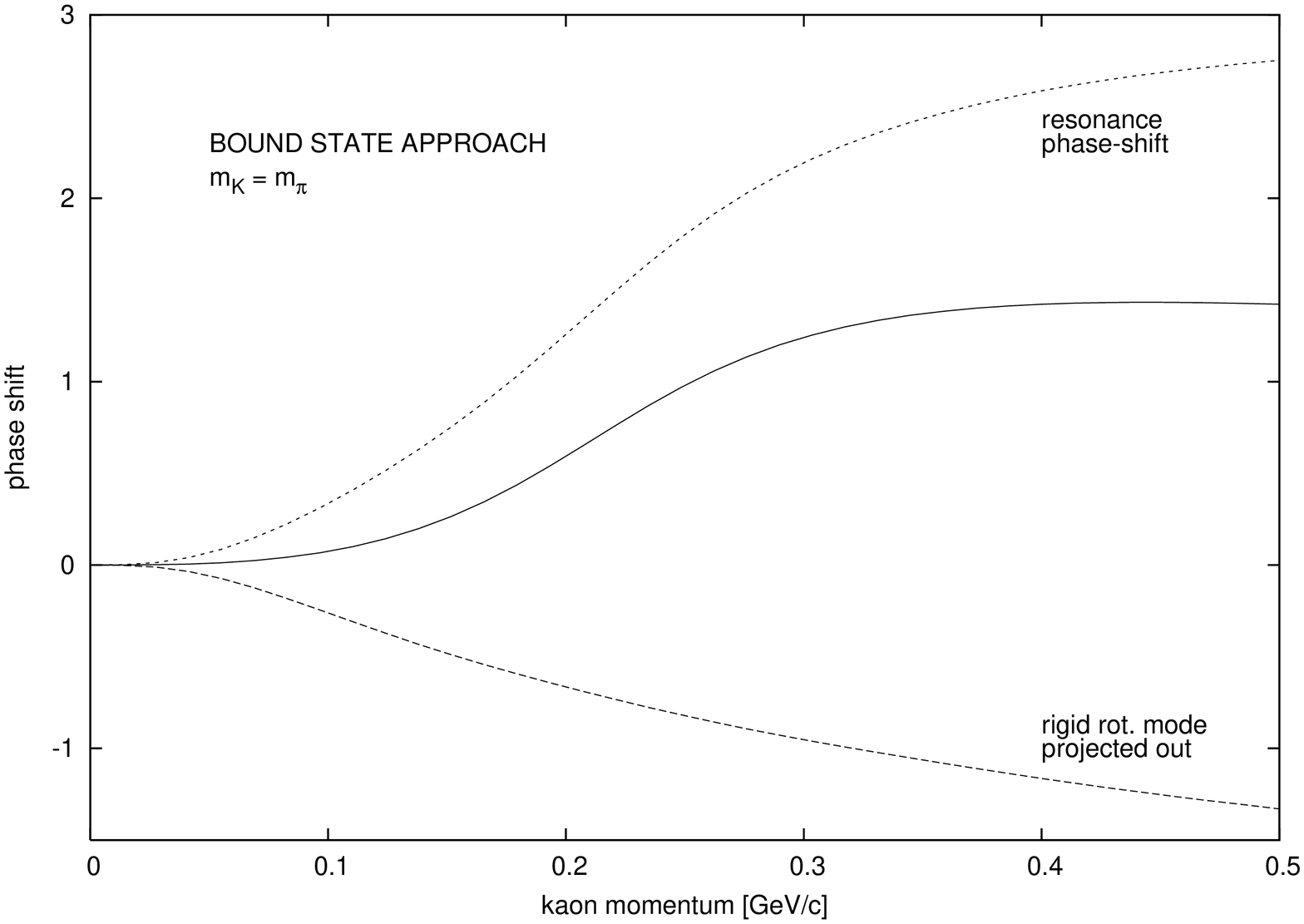,width=5cm,height=6cm,angle=270}\hspace{2cm}
\epsfig{file=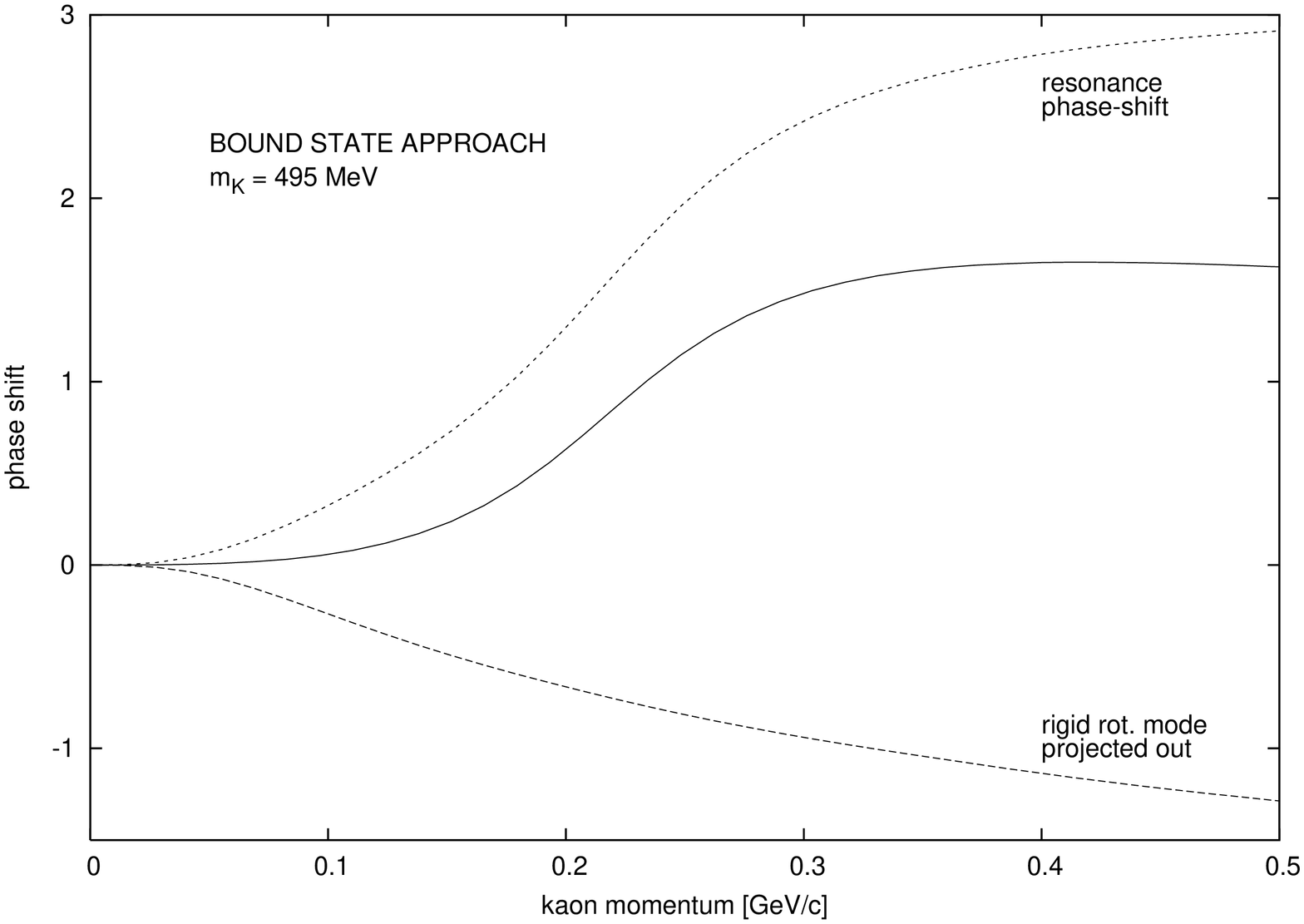,width=5cm,height=6cm,angle=270}}
\caption{\label{fig_2}The full lines are the same phase shifts as in 
figure~\ref{fig_1}. The dashed lines are the background phase shifts 
in the strangeness $S=+1$ channel as functions of the kaon momentum 
$k^2=\omega^2-m_K^2$ as computed from the integro--differential 
equation~(\ref{intdiff}). The dotted lines give their difference 
to be identified as the resonance phase shift. Left panel
$m_K=m_\pi$, right panel $m_K\ne m_\pi$.}
\end{figure}

The constraints only concern the P--wave channel and read
\be
\Psi=\int_0^\infty r^2 dr\, z(r) \left[
\overline{\Pi}(r)-\lambda(r)\overline{\eta}(r)\right] = 0 
\qquad{\rm and}\qquad
\chi=\int_0^\infty r^2 dr\, z(r) M_K(r) \overline{\eta}(r) =0\,.
\label{constraints}
\ee
We add these constraints to the Hamiltonian with global 
Lagrange--multipliers $\alpha$ and $\beta$. From this we obtain the 
homogeneous equations of motion for the fluctuations $\overline{\Pi}$ 
and $\overline{\eta}$, that are orthogonal to the rotations of the soliton, 
as modifications to eqs.~(\ref{blaizoteqm})
\bea
\overline{\Pi}(r)&=&
\left[\omega M_K(r)-\lambda(r)\right]\overline{\eta}(r)
+\alpha M_K(r) z(r)\,,\cr\cr
h^2\,\overline{\eta}(r)+\frac{\lambda^2(r)}{M_K(r)}&=&
\left[\omega -\frac{\lambda(r)}{M_K(r)}\right]\overline{\Pi}(r)
-\alpha\lambda(r)z(r)-\beta M_K(r)z(r)\,.
\label{blaizoteqmconst}
\eea
We multiply the above equations by $z(r)$ from the left and integrate
over the radial coordinate to determine the Lagrange--multipliers 
using the constraints, eq.~(\ref{constraints}) 
\be
\alpha=\int_0^\infty r^2 dr\, z(r)\left[2\lambda(r)\right]\overline{\eta}(r)
\quad{\rm and}\quad
\beta=\omega_0\alpha+\left(m_K^2-m_\pi^2\right)
\int_0^\infty r^2 dr\, z(r)\overline{\eta}(r)\,,
\label{lagmulti}
\ee
since $\int r^2dr zM_Kz=1$ and $\int r^2dr z2\lambda z=\omega_0$. We 
eliminate $\overline{\Pi}$ from the second equation~(\ref{blaizoteqmconst}),
substitute the above Lagrange--multipliers and obtain the 
linear second order integro--differential equation for $\overline{\eta}$,
\bea
h^2\,\overline{\eta}(r)+
\omega\left[2\lambda(r)-\omega M_K(r)\right]\overline{\eta}(r)&=&
-\left[2\lambda(r)-\left(\omega+\omega_0\right)M_K(r)\right]z(r)
\int_0^\infty r^{\prime2}dr^\prime z(r^\prime)
2\lambda(r^\prime)\overline{\eta}(r^\prime)\cr\cr
&&+\left(m_K^2-m_\pi^2\right) \int_0^\infty r^{\prime2}dr^\prime 
z(r^\prime) \overline{\eta}(r^\prime)\,.
\label{intdiff}
\eea
Any solution $\overline{\eta}$ to this equation automatically 
satisfies $\int r^2dr zM_K\overline{\eta}=0$. This is easily 
confirmed by multiplying eq.~(\ref{intdiff}) from the left with
$z(r)$ and integrating over the radial coordinate. Obviously
this integro--differential equation projects eq.~(\ref{scndorder})
on the subspace orthogonal to the rigid rotations of the hedgehog.
As a result the integro--differential equation 
does \emph{not} have a bound state solution to be associated 
with the $\Lambda$ hyperon. In figure~\ref{fig_2} the dashed
lines are the phase shifts $\overline{\delta}$ for strangeness $S=+1$
as computed from the integro--differential equation~(\ref{intdiff}). 
This phase shift is quite shallow with an indication of a small
repulsive potential. 
As explained at the beginning of this subsection, $\overline{\eta}$
parameterizes the background part. That is, $\overline{\delta}$ is the 
\emph{background} phase shift in the strangeness $S=+1$ channel. On the 
other hand the total phase shift, $\delta$ ({\it cf.} figure~\ref{fig_1})
is the sum of the \emph{resonance} phase shift, $\delta_R$ and the 
background phase shift,
\be
\delta_R(k)=\delta(k)-\overline{\delta}(k)\,.
\label{resps}
\ee
We also display this resonance phase shift in figure~\ref{fig_2}.
Obviously it exhibits a clear resonance structure as $\delta_R(k)$ 
smoothly passes through $\pi/2$. Thus we reach a central result
of our studies: Once the background phase shift is properly identified
and subtracted, chiral soliton models indeed predict a distinct resonance 
structure in the strangeness $S=+1$ channel even and in particular 
for $N_C\to\infty$. In section V we will show that in the 
limit $N_C\to\infty$ it is exactly $\delta_R$ that arises 
from exchange of the $\Theta^+$ pentaquark (and the $\Lambda$ hyperon).
Figure~\ref{fig_2} suggests that the resonance is quite broad.
In section VI we will discuss that finite $N_C$ effects are substantial
in sharpening this resonance. Note also from figure~\ref{fig_2} that
the position at which $\delta_R(k)$ passes $\pi/2$ is somewhat lower
than predicted by the estimate, eq.~(\ref{omt}). In section VI we 
will explain that this pole shift originates from the well--know 
effect of iterated resonance exchanges. Actually this is standard in 
scattering theory and does not alter the observation that $\omega_\Theta$
is the "bare" resonance or pole mass.

\subsection{Hyperfine splitting}

Having established the existence of states in both the $S=-1$
(as bound state) and $S=+1$ (as resonance) channels we would like 
to complete this section discussing the hyperfine splitting 
in the BSA for later comparison with the RRA. Even more, we have 
seen that the corresponding kaon wave--functions are dominated by 
the zero mode component, $z(r)$. {\it A posteriori} this justifies 
the estimates for the mass differences to the nucleon, eqs.~(\ref{oml}) 
and~(\ref{omt}). These estimates concern the large--$N_C$ limit 
and are thus degenerate in spin and isospin. To lift this degeneracy, 
states with good spin and isospin need to be constructed in the framework of 
the BSA. For the $S=-1$ channel this has been performed in ref.~\cite{Kl89}
and we will generalize it to also apply for $S=+1$. Starting point
is the parameterization, eq.~(\ref{Usu2}) with the zero mode wave--function
substituted for $\eta_\alpha$ in the $P$--wave, {\it cf.} eq.~(\ref{pwave}).
Spin and isospin states are then generated by canonical quantization
of the $SU(2)$ collective coordinates, $A_2$. Substituting this 
parameterization into the action adds
\be
\Delta L=\frac{\Theta_\pi}{2\,}\, \vec{\Omega\,}^2
+i\left(2\Theta_K-\Theta_\pi\right)
\vec{\Omega\,} \cdot
\left[\left(\frac{dK(t)}{dt}\right)^\dagger\vec{\tau\,}K(t)-
K^\dagger(t)\vec{\tau\,}\frac{dK(t)}{dt}\right]
-\frac{N_C}{2}\vec{\Omega\,}\cdot K^\dagger(t)\vec{\tau\,}K(t)
\label{hyper1}
\ee
to the Lagrange function~\cite{Kl89}. Here 
$\sqrt{2}K(t)=\int d\omega \chi(\omega){\rm e}^{i\omega t}$ 
contains the amplitudes for the kaonic states in the $S=\pm1$ channels.
The dependence on the collective coordinates is
parameterized by the angular velocities
\be
A_2^\dagger(t)\, \frac{d A_2(t)}{dt} 
= \frac{i}{2}\, \vec{\tau}\cdot\vec{\Omega}\,.
\label{su2omega}
\ee
In addition the moment of inertia for spatial rotations~\cite{Ad83} appears,
\be
\Theta_\pi=\frac{2}{3}\int d^3r\, {\rm sin}^2F \left[f_\pi^2
+\frac{1}{\epsilon^2}
\left(F^{\prime2}+\frac{{\rm sin}^2F}{r^2}\right)\right]
=5.115{\rm GeV}^{-1}\, \frac{N_C}{3}\,.
\label{thetapi}
\ee
Treating this additional term perturbatively modifies the 
Hamiltonian by
\be
\Delta H=\frac{1}{2\Theta_\pi}\left[\vec{J\,}
-\frac{i}{2}\left(1-\frac{\Theta_\pi}{2\Theta_K}\right)
\left(\Pi^\dagger\vec{\tau\,}K-K^\dagger\vec{\tau\,}\Pi\right)
+\frac{N_C\Theta_\pi}{4\Theta_K}
K^\dagger\vec{\tau\,}K\right]^2\,,
\label{hyper2}
\ee
where $J_i=\frac{\partial\Delta L}{\partial \Omega_i}$ and 
$\Pi$ is conjugate to $K^\dagger$. The computation of matrix
elements of $\Delta H$ is standard~\cite{Kl89} and yields 
the \emph{hyperfine} contributions to the masses of baryons with 
strangeness $S=\pm1$,
\be
\Delta M_S = \frac{1}{2\Theta_\pi}\left[
c_SJ\left(J+1\right)+\left(1-c_S\right)I\left(I+1\right)
+\frac{3}{4}c_S\left(c_S-1\right)\right]\,,
\label{MBSA}
\ee
with $J$ and $I$ being the spin and isospin quantum numbers
of the considered baryon, respectively. The hyperfine 
parameters
\be
c_{-}=1-\frac{4\Theta_\pi\omega_\Lambda}{8\Theta_K\omega_\Lambda+N_C}
\qquad {\rm and} \qquad
c_{+}=1-\frac{4\Theta_\pi\omega_\Theta}{8\Theta_K\omega_\Theta-N_C}\,.
\label{cpara}
\ee
are ${\cal O}(N_C^0)$ so that the
coefficient $1/\Theta_\pi$ in eq.~(\ref{MBSA}) ensures 
$\lim_{N_C\to\infty}\Delta M_S=0$. We remark that the expressions 
in eqs.~(\ref{oml}), (\ref{omt}) and (\ref{cpara}) are the BSA
predictions for fluctuations, $\eta$ in the subspace of rigid
rotations.

To this end, the mass differences of the $\Lambda$--hyperon and the
$\Theta^+$ pentaquark with respect to the nucleon read
\be
M_{\Lambda,\Theta}-M_N=\omega_{\Lambda,\Theta}
+\Delta M_S -\frac{3}{8\Theta_\pi}
\label{DeltaM}
\ee
Similar relations can be easily written down for other $S=\pm1$ baryon 
states. The last term in eq.~(\ref{MBSA}) is often omitted as it contains
terms that are quartic in the fluctuating field and the calculation
is not completely carried out at this order. The comparison with 
the RRA suggests that these quartic terms contribute 
$9/8\Theta_K$ to the mass of the $S=1$ baryons. This is not 
negligible for $N_C=3$ and clearly shows that subleading contributions
\emph{must} be considered for predictions to be reliable. These 
subleading contributions can be evaluated within the RRA that we 
will discuss in the next section.

\section{Rigid rotator approach (RRA)} 

In this section we will explain how baryon states emerge as 
soliton excitations that are members of flavor $SU(3)$
representations. They are obtained from canonical quantization of the 
collective coordinates $A_3$ in the {\it ansatz}, eq.~(\ref{Usu3}).
For this purpose we do not need to consider the fluctuations,
$\widetilde{\eta}_\alpha$ and relegate the discussion of the 
coupling between $A_3$ and $\widetilde{\eta}_\alpha$ to the 
following section. The resulting Lagrangian for the collective 
coordinates reads~\cite{Gu84}
\be
L=-M+\frac{1}{2}\Theta_{\pi} \Omega_i^2 +
\frac{1}{2}\Theta_K \Omega_\alpha^2 - 
\frac{N_C}{2\sqrt 3}\,\Omega_8-\frac{1}{2}\Gamma\left(1-D_{88}\right)\,.
\label{lcol}
\ee
The soliton mass, $M$ is the functional which upon minimization 
determines the chiral angle, $F(r)$. Its particular form~\cite{Ad83} 
is of no relevance here because we will only consider mass differences.
The moments of inertia, $\Theta_\pi,\Theta_K$ and the symmetry breaking 
parameter, $\Gamma$ have already been introduced in the previous 
section, {\it cf.} eqs.~(\ref{zeromode}),~(\ref{omscnd}) and~(\ref{thetapi}).
In equation~(\ref{lcol}) also angular velocities for rotations 
into strangeness direction appear. These are straightforward 
generalizations of the definition~(\ref{su2omega}) to $A_3$
\be
A_3^\dagger(t)\, \frac{d A_3(t)}{dt}
= \frac{i}{2}\, \lambda_a\Omega_a\,.
\label{su3omega}
\ee 
Finally the symmetry breaking part of the action, eq.~(\ref{SB}) 
causes the explicit dependence on the collective coordinates. This
is conveniently displayed with the help of their adjoint representation,
\be
D_{ab}(A_3)=\frac{1}{2}\,{\rm tr}\,\left[
\lambda_a A_3 \lambda_b A_3^\dagger\right]\,.
\label{Dab}
\ee
We note that in more complicated soliton models ({\it e.g.\@} with 
vector mesons or chiral quarks) the symmetry breaking part of $L$ 
has a richer structure~\cite{We96} which eventually leads to better
agreement between predicted and observed baryon spectra. 
However, we presently compare two different
treatments (BSA vs. RRA) of the \emph{same} model and thus other
symmetry breaking terms are of no relevance to us. For quantization
we compute the conjugate momenta, 
\be
R_a=-\frac{\partial L}{\partial\Omega_a}=
\begin{cases}
-\Theta_\pi^2\Omega_i=-J_i\,, &a=i=1,2,3\cr
-\Theta_K\Omega_\alpha\,, &a=\alpha=4,..,7\cr
\frac{N_C}{2\sqrt3}\,, &a=8
\end{cases} \quad .
\label{Rgen}
\ee
The quantization prescription then demands the commutation
relations $[R_a,R_b]=-if_{abc}R_c$ with $f_{abc}$ being the
antisymmetric structure constants of $SU(3)$. That is, the $R_a$ are 
the generators of the intrinsic $SU(3)$ group. The hedgehog 
structure of the classical soliton ensures that $J_i=-R_i$, 
($i=1,2,3$) is the total spin operator. The Wess--Zumino
term, eq.~(\ref{WZ}) plays a decisive role. It is non--local
and only linear in the time--derivative such that it contributes
the term linear in $\Omega_8$ to the collective coordinate 
Lagrangian. Therefore the corresponding conjugate momentum
is constrained to $R_8=\frac{N_C}{2\sqrt3}$ and we require 
a Lagrange multiplier, $\mu$ for this constraint when constructing
the Hamiltonian for the collective coordinates by Legendre 
transformation from eq.~(\ref{lcol})
\be
H = M +\frac{1}{2\Theta_\pi}  R_i^2
+ \frac{1}{2\Theta_K}  R_\alpha^2
+ \frac{1}{2}\Gamma \big( 1 - D_{88} \big)
- \mu \left(R_8-\frac{N_C}{2 \sqrt{3}} \right)\,.
\label{hcol}
\ee
Demanding that $L$ can be obtained from $H$ by the inverse Legendre 
transformation requires to set $\mu=\Omega_8$. Since $[H,R_8]=0$ 
this angular velocity remains undetermined and different choices merely 
imply different gauges. This vanishing commutator also suggests that 
the constraint can be implemented on the eigenstates of $H$. Some of 
the discussion on finding these eigenstates for arbitrary $N_C$ is 
already contained in the literature~\cite{Kl89}. Here 
we focus on the comparison between BSA and RRA for the $S=+1$ pentaquarks,
which is new. So is the non--perturbative investigation of symmetry breaking 
effects in the RRA for arbitrary $N_C$\footnote{A simplified treatment
of symmetry breaking to only leading order in perturbation theory
gives results consistent with ours~\cite{Ko05}.}.

We concentrate on the case of vanishing symmetry breaking, $\Gamma=0$
for a first discussion on finding the eigenstates of the collective 
coordinate Hamiltonian, eq.~(\ref{hcol}). For $\Gamma=0$ the eigenstates
of $H$ are members of definite irreducible representations of 
$SU(3)$ that are characterized by the labels $(p,q)$. 
The eigenvalues of $H$ can then be computed from the eigenvalue
of the quadratic Casimir operator of $SU(3)$,
\be
C_2=\big\langle R_a^2\big\rangle=
\frac{1}{3}\left(p^2+pq+q^2\right)+p+q\,.
\label{c2ev}
\ee
The constraint on $R_8$ only allows representations with
\be
p+2q=N_C+3t\,
\label{pqconst}
\ee
where $t=0,1,2,\ldots$ refers to Biedenharn's triality~\cite{Bi84}.
The physical hypercharge for baryons is generalized to $Y=\frac{N_C}{3}+S$.
Note that because of $Y_{\rm max}=\frac{N_C}{3}+t$ this triality
equals the largest strangeness quantum number $S_{\rm max}=t$
contained in the representation $(p,q)$. The eigenstates of $H$ are 
most conveniently arranged according to increasing $t$. For $t=0$ 
they are
\be
\left(p,q\right)=\left(2J,\frac{N_C-2J}{2}\right)
\label{pq0}
\ee
where $J=\frac{1}{2},\frac{3}{2},\ldots,\frac{N_C}{2}$ are the 
allowed spin eigenvalues, {\it i.e.\@} $\langle R_i^2\rangle=J(J+1)$. 
{}From equation~(\ref{c2ev}) we compute the corresponding energies
\be
E_0=M+\frac{J(J+1)}{2\Theta_\pi}+\frac{N_C}{4\Theta_K}\,,
\label{e0}
\ee
which coincides with the $SU(2)$ mass formula~\cite{Ad83} up to an 
additive constant for nucleon and $\Delta$ type states. For $t=1$ we find
\bea
\left(p,q\right)&=&\left(2J\mp1,\frac{N_C+3}{2}-\frac{2J\mp1}{2}\right) \cr\cr
E_{\mp}&=&M+\frac{J(J+1)}{2\Theta_\pi}
+\frac{2N_C+4\mp(2J+1)}{4\Theta_K}\,,
\label{epm}
\eea
with $J=\frac{1}{2},\frac{3}{2},\ldots,\frac{N_C}{2}+1$. 
In particular the lowest lying multiplets with spin
$J=\frac{1}{2}$ and $J=\frac{3}{2}$ are
\bea
J=\frac{1}{2}&:& \left(p,q\right)=\left(1,\frac{N_C-1}{2}\right),\,
\left(0,\frac{N_C+3}{2}\right),\,\ldots =
"\mathbf{8}","\mathbf{\overline{10}}",\ldots \cr \cr
J=\frac{3}{2}&:& \left(p,q\right)=\left(3,\frac{N_C-3}{2}\right),\,
\left(2,\frac{N_C+1}{2}\right),\,\ldots =
"\mathbf{10}","\mathbf{27}",\ldots  \,,
\label{lowestreps}
\eea
where the expressions to the furthest right refer to the dimensionalities
of the representations for $N_C=3$. 
The above assignments are unique in soliton models and so are
the flavor quantum numbers attributed to the lowest lying baryon states
within the multiplets. Essentially this is inferred from the 
fact that these assignments minimize the energy functional.  For 
example, for large--$N_C$ the symmetry breaking term contributes 
\be
\frac{\Gamma}{2}\big\langle 1-D_{88}\big\rangle 
\to \frac{\Gamma}{2}\left(1-\frac{3}{N_C}Y\right)=
-\frac{3\Gamma}{2N_C}S \qquad (S\le t)
\label{sbnc}
\ee
to the energy functional if symmetry breaking was treated in
first order perturbation.
For finite $N_C$ this generalizes to the fact
that symmetry breaking causes the masses of baryons within a 
multiplet to decrease with increasing
strangeness. For $t=0$ we therefore obtain the lightest 
$J=\frac{1}{2}$ baryon, the "nucleon" with zero strangeness and
isospin $I=\frac{1}{2}$ and the lightest $J=\frac{3}{2}$ baryon,
the "$\Delta$" with zero strangeness and isospin $I=\frac{3}{2}$.
For $t=1$ the lowest lying "pentaquarks" with $J=\frac{1}{2}$ 
and $J=\frac{3}{2}$ carry strangeness $S=+1$ and isospin $I=0$ and 
$I=1$, respectively. Of course, for finite symmetry breaking mixing 
with higher dimensional multiplets occurs (see below). Although such
non--linear effects may not be omitted, the spin--flavor assignments 
are unaffected.

The lowest lying pentaquarks with $S=+1$ dwell in $t=1$ representations 
and the two mass formulas~(\ref{epm}) refer to the spin--isospin relations
$I=J\mp\frac{1}{2}$. It is straightforward to derive the corresponding
energies
\be
E_{\mp}=M+\frac{J(J+1)}{2\Theta_\pi}+\frac{N_C}{2\Theta_K}
+\frac{1}{2\Theta_K}\left[I(I+1)-J(J+1)+\frac{9}{4}\right]\,.
\label{epm1}
\ee
The terms ${\cal O}(1/N_C)$ exactly match\footnote{Without symmetry breaking
we have $\omega_\Theta=N_C/4\Theta_K$ and thus $1-c_+=\Theta_\pi/\Theta_K$.}
those in eq.~(\ref{MBSA}) when equating the hyperfine energies that are 
related to expressions quartic in the fluctuations to $\frac{9}{8\Theta_K}$.
Even more, we may compute the mass difference between the nucleon
(in $"\mathbf{8}"$) and pentaquarks with $I=0$ and $J=\frac{1}{2}$
(in $"\mathbf{\overline{10}}"$)
\be
E_{\mathbf{\overline{10}}}-E_\mathbf{8}=\frac{N_C+3}{4\Theta_K}\,,
\label{Ntheta}
\ee
which coincides with $\omega_\Theta$ as $N_C\to\infty$, as it should. 
Thus we conclude that the BSA and RRA are consistent when 
flavor symmetry breaking is omitted.
Note that this mass difference acquires a factor 2 for $N_C=3$.

We will next consider the physical case of non--zero flavor 
symmetry breaking.
As a first measure a perturbative treatment of flavor symmetry breaking 
seems tempting because it reproduces the Gell-Mann--Okubo mass 
relations~\cite{Ok62,Ge64} in the RRA for $N_C=3$~\cite{Gu84}. However, 
we already argued in the context of eqs.~(\ref{oml}) and~(\ref{omt}) 
that the perturbation series in $\Gamma$ does not converge for the BSA. 
We recall that many of the pentaquark studies in the RRA treated flavor 
symmetry breaking perturbatively (at best to second 
order)~\cite{Di97,El04}\footnote{In the case of pentaquarks symmetry 
breaking was treated exactly in refs.~\cite{Wa92,Wa92a,We98,Wa03}.}.
This seems at odds with the BSA. Also, the Gell-Mann--Okubo relations are 
not reproduced in the BSA~\cite{We96}. Hence we have to solve the 
eigenvalue problem
\be
\left\{M+\frac{J(J+1)}{2\Theta_\pi}+\frac{R_\alpha^2}{2\Theta_K}
+\frac{1}{2}\Gamma\left(1-D_{88}\right)\right\}\Psi 
=E\Psi
\quad {\rm together~with}\quad 
R_8\Psi=\frac{N_C}{2\sqrt3}\Psi
\label{evprob}
\ee 
(numerically) exactly. For $N_C=3$ this has been accomplished some time 
ago~\cite{Ya88} and frequently used since then for the computation
of a variety of baryon properties,
{\it cf.\@} refs.~\cite{Pa89a,Pa89b,Wa03}.
This can straightforwardly be generalized 
by putting $Y_R=2R_8/\sqrt{3}=N_C/3$ in eq.~(2.18) of 
ref.~\cite{Ya88},  together with the $N_C$ dependencies of 
$\Theta_\pi$, {\it cf.} eq.~(\ref{zeromode}) and $\Gamma$ (paragraph 
after eq.~(\ref{oml})). Equivalently, the operator $D_{88}$ may be
diagonalized in the space of $SU(3)$ representations with 
$R_8\Psi=\frac{N_C}{2\sqrt3}\Psi$ using $SU(3)$ Clebsch--Gordan
coefficients. The only condition on $N_C$ is that it must
be taken odd. In section VII and VIII we will employ the 
resulting wave--functions to compute matrix elements of collective 
coordinate operators for arbitrary odd $N_C$ and non--zero symmetry
breaking. 

\begin{figure}[t]
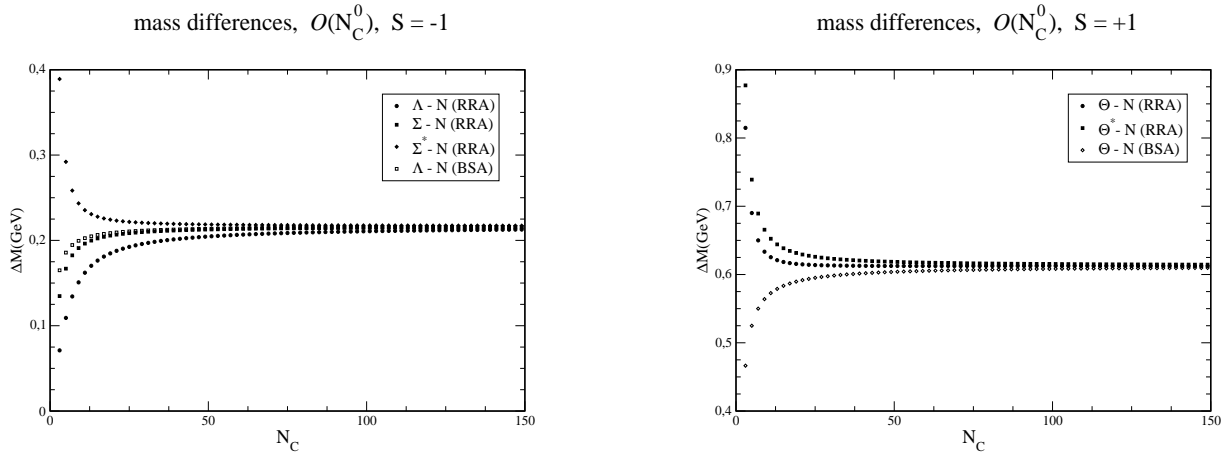

\centerline{\hskip -0.0cm
\epsfig{figure=mass_0m.eps,height=6.0cm,width=7.0cm}\hspace{2cm}
\epsfig{figure=mass_0p.eps,height=6.0cm,width=7.0cm}}
\caption{\label{fig_3}Mass differences at ${\cal O}(N_C^0)$ computed for 
$\epsilon=4.25$ in the Skyrme model as functions of $N_C$. In the 
RRA they are the corresponding differences of the eigenvalues 
in eq.~(\ref{evprob}) while in the BSA they are extracted from
eq.~(\ref{DeltaM}). Left panel $\Delta S=-1$; right panel
$\Delta S=+1$.}
\end{figure}

\begin{figure}[t]
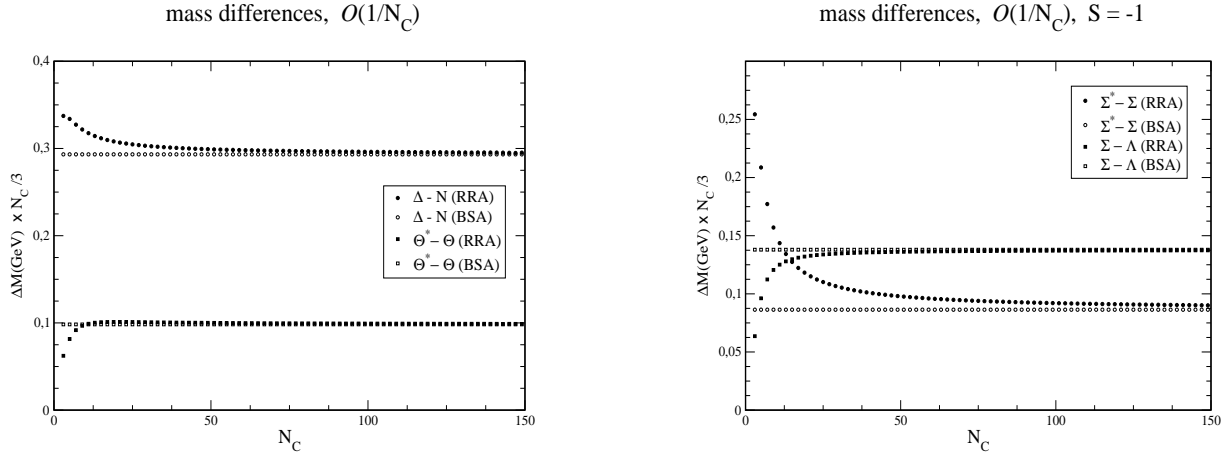

\centerline{\hskip -0.0cm
\epsfig{figure=mass_1p.eps,height=6.0cm,width=7.0cm}\hspace{2cm}
\epsfig{figure=mass_1m.eps,height=6.0cm,width=7.0cm}}
\caption{\label{fig_4}Mass differences at ${\cal O}(1/N_C)$ computed for
$\epsilon=4.25$ in the Skyrme model as functions of $N_C$.
Left panel: baryons with strangeness $S=0,1$; right panel $S=-1$.
See also caption of figure~\ref{fig_3}.}
\end{figure}

For the previously mentioned parameters we present the 
resulting mass differences in figures~\ref{fig_3} and~\ref{fig_4}.
In figure~\ref{fig_3} we concentrate on mass differences that
scale like ${\cal O}(N_C^0)$, {\it i.e.\@} between baryons whose
strangeness quantum numbers differ by one unit. 
As $N_C\to\infty$ the BSA for fluctuations in the rotational subspace
predicts $\omega_\Lambda$ and $\omega_\Theta$
for those mass differences regardless of spin and isospin. Obviously this 
is matched by the RRA. In figure~\ref{fig_4} we display the mass 
differences that scale like ${\cal O}(1/N_C)$, {\it i.e.\@} we compute 
the hyperfine splitting in both approaches. In doing so we only consider
combinations for which in the BSA the ordering ambiguities and omissions
from ${\cal O}(\eta_\alpha^4)$ terms cancels, {\it i.e.\@} baryons 
with identical strangeness.
Again, perfect agreement between the two approaches is observed 
for $N_C\to\infty$. However, at $N_C=3$ sizable $1/N_C$ corrections
occur that cannot be accounted for by the BSA.

A couple of further comments are in order. First, when computing matrix
elements we observe that, {\it cf.\@} eq.~(\ref{sbnc}), 
\be
\frac{\Gamma}{2}\big\langle 1-D_{88}\big\rangle={\cal O}(N_C^0)\,,
\label{sbnc1}
\ee
that is, the net effect of flavor symmetry breaking in the RRA is an 
${\cal O}(N_C^0)$ contribution to the baryon mass. Again, this is consistent 
with the BSA where flavor symmetry breaking enters the ${\cal O}(N_C^0)$ 
equation of motion for the kaon fluctuations~(\ref{scndorder}). Second,
the literature contains claims that the $\Delta$--nucleon
mass difference would scale like ${\cal O}(N_C^0)$ once symmetry
breaking is incorporated in the RRA~\cite{Pr03}. The actual calculation,
displayed in fig~\ref{fig_4}, shows that this is not correct. As in two
extreme cases, zero and infinite\footnote{This is the $SU(2)$ version
of the chiral soliton model.} symmetry breaking, this mass difference
scales like ${\cal O}(1/N_C)$. The physical value, $m_K=495{\rm MeV}$
is intermediate to these extreme cases and it thus comes to no 
surprise that this mass difference vanishes as $N_C\to\infty$.

The second remark concerns the eventual restriction
of collective coordinates to only dynamical zero modes.
For zero symmetry breaking ($\Gamma=0$) the rigid rotation 
eigenstates of eq.(\ref{evprob}) are states in the 
$\mathbf{8}$, $\mathbf{10}$, $\overline{\mathbf{10}}$, 
$\mathbf{27}$,...  dimensional representations of flavor $SU(3)$.
Though the anti--kaon fluctuations are zero frequency solutions
in the BSA, the kaon fluctuations are not due to the Wess--Zumino 
term, {\it i.e.\@} the latter are not really dynamical zero--modes. 
Therefore it has been suggested to introduce collective coordinates 
\emph{only} for the anti--kaonic zero--modes in a manner that 
only the octet and decuplet would
be treated collectively~\cite{Ch05}. Such an approach seems 
complicated because constraints on both the fluctuations and the 
collective coordinates are required. Putting the question of 
feasibility aside, there is no advantage in doing so
to study the $\Theta^+$ pentaquark since such a treatment 
essentially corresponds to the BSA and prevents the incorporation
of finite $N_C$ effects. The restriction to the octet and decuplet
seems even more problematic in view of flavor symmetry breaking 
($\Gamma\ne0$) as it mixes baryon states from different $SU(3)$ 
representations~\cite{Pa89a}. In particular the nucleon wave--function 
contains admixture of the nucleon type state in the 
$\overline{\mathbf{10}}$. Thus the latter may not be omitted.

To round up this section on the RRA we note that other
approaches to quantize chiral solitons in flavor $SU(3)$, like the
slow rotator~\cite{Schw92} or the breathing mode~\cite{Sch91a} approaches,
are essentially extensions of the RRA.

So far we have established the relation between rotational excitations
of the soliton and the bound states in the large--$N_C$ limit. In order
to study the decay width of the $\Theta^+$ pentaquark we need to
introduce fluctuations by extending the RRA to the RVA. 

\section{Collective rotations and vibrations}

We will now include the kaon fluctuations in the RVA according to
the {\it ansatz}~(\ref{Usu3}) and compare it to results for the 
fluctuations in the BSA, section III. The flavor rotating hedgehog
$A_3(t)U_0(\vec{x\,})A^\dagger_3(t)$ is \emph{not} a solution
to the field equations. Hence the expansion of the action 
with respect to $\widetilde{\eta}$ has a linear term. This linear
term gives rise to Yukawa couplings like $KN\Lambda$ and 
$KN\Theta$. As a consequence the $KN$ phase shift in the isospin
$I=0$ channel\footnote{The $I=1$ channel, which we discuss in the
last paragraph of section VIII, has $\Sigma^*$ and $\Theta^*$ exchange 
contributions.} has contributions from (iterated) exchange of 
$\Lambda$ and $\Theta$. Technically, these exchange contributions 
emerge in the equation of motion for $\widetilde{\eta}$ as a separable 
potential of the form
"(Yukawa interaction)$\times$(propagator)$\times$(Yukawa interaction)".
The reader may want to peak ahead to eq.~(\ref{Yukawa1}) for
the explicit expression. We have already noted 
that the background field $\overline{\eta}$, that we have computed in 
section III, obeys an equation of motion that is obtained from the one
for $\widetilde{\eta}$ by omitting the Yukawa couplings. That is,
the respective equations of motion differ by this separable potential
and we can straightforwardly construct $\widetilde{\eta}$ from 
$\overline{\eta}$ using standard Lippmann--Schwinger equation techniques. 
This will make the role of exchanged resonances very transparent. For 
convenience we list the notation for the involved radial functions
$$
\begin{array}[b]{ll}
\eta(r):& \mbox{\rm solution to eq.~(\ref{scndorder}) in the BSA} \cr
\overline{\eta}(r):& \mbox{\rm solution to the integro--differential
equation,~(\ref{intdiff}) with constraints} \cr
\widetilde{\eta}(r):& \mbox{\rm solution to the integro--differential
equation,~(\ref{Yukawa1}) with constraints and Yukawa exchange}
\end{array}
$$

We will establish two central results in the large--$N_C$ limit:
\begin{itemize}
\item[A)]
We compute the RVA fluctuations $\widetilde{\eta}$ by 
supplementing the equations of motion for $\overline{\eta}$ with the 
effects due to $\Lambda$ and $\Theta$ exchanges. We compute the phase 
shifts for $\widetilde{\eta}$ and show that they are \emph{identical} 
to those computed from $\eta$ in the BSA, eq.~(\ref{Usu2}).
\item[B)]
We separate the Yukawa exchange contributions to the scattering 
amplitude and find the corresponding phase shift to \emph{equal} 
the resonance phase shift displayed in figure~\ref{fig_2}.
\end{itemize}
In doing so we need to identify the Yukawa couplings by computing
the terms linear in $\widetilde{\eta}$. Note that in the action
all linear terms must occur as iso--singlets. Then there are two 
sources for Yukawa terms: 1)~the flavor symmetric piece 
$\Gamma_{SK}+\Gamma_{WZ}$ has couplings to the angular velocities 
$\Omega_a$, 2)~the symmetry breaking piece has a contraction between 
$D_{8a}$ and $\widetilde{\eta}_a$. For the moment we will only consider the 
first case and relegate the discussion of flavor symmetry breaking
to section VII.

In the flavor symmetric case there are already many rotation--vibration 
coupling terms linear and quadratic in $\Omega_a$. In order to keep the 
calculation feasible we keep only those terms that survive in the 
large--$N_C$ limit. The so--selected Yukawa coupling is
then treated to \emph{all} orders in~$N_C$. Admittedly, this procedure 
is not completely consistent and the neglected couplings are
not necessarily small at $N_C=3$. However, a major purpose of
the present investigation is to show the equivalence with the BSA.
For this purpose it completely suffices to only consider the leading $N_C$ 
couplings. In order to isolate these terms 
we need to discuss the $N_C$ scaling of the angular velocities
$\Omega_a$ to identify those Yukawa terms that survive in the 
large--$N_C$ limit. That is straightforward for $a=1,\ldots,7$ as
we simply invert eq.~(\ref{Rgen}),
\bea
\sum_{i=1}^3\Omega_i^2&=&\frac{J(J+1)}{\Theta_\pi^2}
\quad{\Rightarrow}\quad \Omega_i={\cal O}\left(1/N_C\right)
\quad \mbox{\it i.e.\@}\quad R_i={\cal O}(N_C^0) \,,
\cr \cr
\sum_{\alpha=4}^7\Omega_\alpha^2&=&
\frac{C_2-J(J+1)-N_C^2/12}{\Theta_K^2}
\quad{\Rightarrow}\quad \Omega_\alpha={\cal O}\left(1/\sqrt{N_C}\right)
\quad \mbox{\it i.e.\@}\quad R_\alpha={\cal O}(\sqrt{N_C}) \,,
\label{ordOmega}
\eea
as the moments of inertia $\Theta_\pi$ and $\Theta_K$ are 
${\cal O}\left(N_C\right)$. In the large--$N_C$ expansion 
$\Omega_i$ ($i=1,2,3$) is thus subleading to $\Omega_\alpha$
($\alpha=4,\ldots,7$) and may be omitted. 
As shown in the appendix, the angular velocity $\Omega_8$ may be 
absorbed in a suitable redefinition of the energy scale
\be
\omega\longrightarrow\omega+\frac{\sqrt3}{2}\Omega_8\,,
\label{omega8}
\ee
where $\omega$ is the frequency of the kaon fluctuation, {\it cf.} 
eq.~(\ref{pwave}). Thus, to compute the $\Omega$--$\widetilde{\eta}$ 
Yukawa coupling we only need to consider the kaonic angular 
velocities $\Omega_\alpha=-R_\alpha/\Theta_K$ with 
$\alpha=4,\ldots,7$. This lengthy calculation described in the
appendix not only requires to extract the terms in the action 
that are linear in $\widetilde{\eta}$ but also to consider the 
constraints, eq.~(\ref{constraints}) because they contain pieces
that are of the same linear order in $\widetilde{\eta}$. 
Having extracted these linear terms we are unfortunately 
not yet in the position to compute matrix elements 
of the interaction Hamiltonian~(\ref{hint}) between $\Theta$ and 
$KN$ states in the space of the collective coordinates.
The reason is that $\widetilde{\eta}$ parameterizes the kaon field in 
the intrinsic frame. A collective flavor rotation relates
the intrinsic fluctuations to those in the laboratory frame, that
are observed in kaon--nucleon scattering,
\be
\widetilde{\xi}_a=D_{ab}\widetilde{\eta}_b\,,
\label{labframe}
\ee
with $a,b=1,\ldots,8$. In the large--$N_C$ limit, $D_{\alpha i}$ and 
$D_{\alpha 8}$, the couplings to the intrinsic pions and $\eta$, 
respectively, are suppressed. Hence we approximate
\be
\widetilde{\xi}_\alpha=D_{\alpha\beta}\widetilde{\eta}_\beta\,,
\label{labframe1}
\ee
with $\alpha,\beta=4,\ldots,7$. 
Finally the resulting interaction Hamiltonian reads,
\be
H_{\rm int}=\frac{2f_\pi}{\Theta_K}\,
d_{i\alpha\beta}\, D_{\gamma\alpha}R_\beta\, \int d^3r\,
{\rm sin}\frac{F(r)}{2}\left[2\lambda(r)-\omega_0 M_K(r)\right]
\hat{x}_i \widetilde{\xi}_\gamma(\vec{x\,},t)R_\beta\,,
\label{hint}
\ee
where $d_{i\alpha\beta}$
are the symmetric structure constants of $SU(3)$. The pieces
involving $\omega_0=N_C/4\Theta_K$ and $\lambda(r)$
stem from $\Gamma_{SK}$ and $\Gamma_{WZ}$, respectively. From
eqs.~(\ref{parameters}),~(\ref{ordOmega}) we easily verify that
$H_{\rm int}={\cal O}(N_C^0)$.

We need to compute matrix elements of the 
interaction Hamiltonian, eq.~(\ref{hint}) between a kaon nucleon 
state coupled to good spin ($J$) and isospin ($I$) and a collective
excitation with the same quantum numbers. We denote the operator that 
annihilates a kaon in the laboratory frame with isospin projection 
$i=\pm\fract{1}{2}$ and orbital angular momentum ($\ell,m$) by
$a^{(i)}_{\ell,m}$ and use the Wigner--Eckart theorem to 
reduce the matrix elements of $H_{\rm int}$ to those computed
in the space of the collective coordinates,
\be
\sum_{i=\pm\frac{1}{2}}
\langle \Theta |a^{(i)}_{1,m}d_{m\alpha\beta}
D_{\gamma(i)\alpha}R_\beta|
(KN)_{I=0,J=\frac{1}{2}}\rangle=\sqrt{6}\,
\langle \Theta^+ \uparrow| d_{3\alpha\beta} 
D_{+\alpha}R_\beta|n \uparrow \rangle
\label{WE1}
\ee
where $\gamma(\pm\fract{1}{2})$ maps the isospin projection 
onto the flavor indices,
$D_{\pm a}=D_{4a}\pm iD_{5a}$. The arrow indicates the 
spin projection $J_3=\fract{1}{2}$. In what follows we will always 
adopt this value without explicit mention. Orbital angular 
momentum $\ell=1$ is projected onto by $\hat{x}_i$ in
$H_{\rm int}$. Of course, that is nothing but the statement
that only the P--wave channel is affected by the rigid rotations. 
A similar reduction formula holds for the matrix element of 
$H_{\rm int}$ between final $\Lambda$ and initial anti--kaon 
nucleon states. In the flavor symmetric case the relevant matrix elements 
of the collective coordinate operators are
\bea
\langle \Theta^+| d_{3\alpha\beta} D_{+\alpha}R_\beta|n\rangle &=&
\frac{N_C+1}{4}\sqrt{\frac{N_C+1}{2\left(N_C+3\right)\left(N_C+7\right)}}
=:X_\Theta\sqrt{\frac{N_C}{32}}\cr\cr
\langle \Lambda| d_{3\alpha\beta}
D_{-\alpha}R_\beta|p\rangle &=& 
-\frac{1}{2}\frac{N_C+5}{\left(N_C+7\right)\sqrt{2N_C+6}}
=:X_\Lambda\sqrt{\frac{N_C}{32}}\,.
\label{ddr}
\eea
For the matrix elements in 
eq.~(\ref{ddr}) we have factored the large--$N_C$ value
for later convenience 
\bea
X_\Theta&=&
\sqrt{\frac{\left(N_C+1\right)^3}{N_C\left(N_C+3\right)\left(N_C+7\right)}}
=\begin{cases}
1\,, & \quad N_C\to\infty \cr
\frac{4}{3\sqrt{5}}\,, & \quad N_C=3
\end{cases} \cr\cr
X_\Lambda&=&-\frac{2\left(N_C+5\right)}
{\left(N_C+7\right)\sqrt{N_C\left(N_C+3\right)}}
=\begin{cases}
0\,, & \quad N_C\to\infty \cr
-\frac{4\sqrt{2}}{15}\,, & \quad N_C=3
\end{cases}\,.
\label{xddr}
\eea
Only the $\Theta^+$ Yukawa coupling survives as $N_C\to\infty$. Hence 
there is only one intermediate state in that limit. For the physical 
value, $N_C=3$, the collective coordinate part of the $\Lambda$ Yukawa 
coupling constant is actually larger in magnitude than that of $\Theta^+$.

Obviously this explicit derivation of the interaction Hamiltonian
yields only a \emph{single} $SU(3)$ 
structure ({\it i.e.\@} $d_{i\alpha\beta}\hat{x}_iD_{a\alpha}R_\beta$)
for the $\Theta-N$ transition. This is to be contrasted 
with earlier studies~\cite{Di97,Pr03,El04,Pr05} who "invented" three 
structures, even in the flavor symmetric case. 

We want to add the particle exchanges induced by $H_{\rm int}$ to the 
differential equation,~(\ref{intdiff}) for the constrained fluctuations,
$\overline{\eta}$ (still omitting symmetry breaking). The propagators 
are $(\omega_\Theta-\omega)^{-1}$ and $(\omega_\Lambda+\omega)^{-1}$ 
for $\Theta$ and $\Lambda$, respectively, because the differential equations 
for kaons and anti--kaons are related by $\omega\leftrightarrow-\omega$.
Thus the Yukawa terms induce the separable potential
\be
\frac{\left|\langle\Theta|H_{\rm int}|(KN)_{I=0}\rangle\right|^2}
{\omega_\Theta-\omega}
+\frac{\left|\langle\Lambda|H_{\rm int}|(KN)_{I=0}\rangle\right|^2}
{\omega_\Lambda+\omega}\,.
\label{potential}
\ee
Note that these matrix elements concern the $T$--matrix elements in the 
laboratory frame. However, we observe from eq.~(\ref{pwaveint})
that for the $\Theta^+$ channel the laboratory and 
intrinsic $T$--matrix elements are identical. Hence we may 
add the exchange potential, eq.~(\ref{potential}) in the intrinsic
frame. We then end up with the modified integro--differential equation
\bea
&&h^2\widetilde{\eta}(r)+
\omega\left[2\lambda(r)-\omega M_K(r)\right]\widetilde{\eta}(r)=
-z(r)\left[\int_0^\infty r^{\prime2}dr^\prime z(r^\prime)
2\lambda(r^\prime)\widetilde{\eta}(r^\prime)\right]\cr
&&\hspace{2cm}\times
\left[2\lambda(r)-\left(\omega+\omega_0\right)M_K(r)
-\omega_0\left(\frac{X_\Theta^2}{\omega_\Theta-\omega}
+\frac{X_\Lambda^2}{\omega}\right)
\left(2\lambda(r)-\omega_0M_K(r)\right)\right]\,,\hspace{1cm}
\label{Yukawa1}
\eea
in the flavor symmetric case. The particular coefficient, $\omega_0$ 
of the Yukawa contribution will become clear when we discuss the
width of $\Theta^+$ in the next section. The Yukawa contribution
to the integro--differential equation is orthogonal to the zero--mode,
\be
\int_0^\infty r^2dr\,
z(r)\left[2\lambda(r)-\omega_0M_K(r)\right]z(r)=0
\label{Yukawa2}
\ee
and thus any solution to eq.~(\ref{Yukawa1}) still satisfies the
second constraint in eq.~(\ref{constraints}) which we subsequently
employ to simplify the integral.

The essential observation is that eq.~(\ref{Yukawa1}) has 
a simple solution as $N_C\to\infty$ ($X_\Theta=1$ and $X_\Lambda=0$)
\be
\widetilde{\eta}(r)=\eta(r)-az(r)
\qquad {\rm with} \qquad
a=\int_0^\infty dr r^2\, z(r) M_K(r)\eta(r)\,.
\label{solution}
\ee
Here $\eta(r)$ is the solution to the unconstrained equation~(\ref{scndorder})
in the BSA. The radial function, $z(r)$ associated with the zero--mode is 
localized in space. Thus the phase shifts of $\eta$ and $\widetilde{\eta}$,
which are extracted from the respective asymptotic behaviors are 
identical. This proves our assertion A). In turn this shows that
the difference of the phase shifts defined in eq.~(\ref{resps}) is 
directly related to the resonance exchange. This proves our assertion B).
In section VII we will show that these assertions also hold when symmetry 
breaking is included even though that case is significantly more involved 
as {\it e.g.\@} $\lim_{N_C\to\infty}X_\Lambda\ne0$.

\section{The width of the $\Theta^+$}

In this section we will show that the Yukawa terms can be imposed to 
compute the width of the exchanged particles. For simplicity we will
still omit flavor symmetry breaking. 

\subsection{Large--$N_C$ limit}

In a first step we show that in the large--$N_C$ limit the iterated exchange, 
that is induced by the Yukawa terms in eq.~(\ref{Yukawa1}), leads to the
resonance phase shift discussed already in section IV. That is, we want 
to prove assertion B) by showing that the resonance width associated 
with $\delta_R(k)$ equals that induced by the Yukawa coupling of 
$\Theta^+$.

For large--$N_C$ we may omit the $\Lambda$--exchange. Then the 
reaction matrix formalism is most suitable to compute $\delta_R(k)$ from 
the iterated exchange of a single resonance. The (real) reaction matrix 
obeys the Lippmann--Schwinger equation
\be
R=V+V\overline{G}_0R\,,
\label{lseq}
\ee
where $\overline{G}_0$ and $V$ are the Green's function for the constrained 
problem, eq.~(\ref{intdiff}) and the separable potential due to the particle 
exchange, eq.~(\ref{potential}), respectively. We stress that this is 
an exact solution to the scattering problem rather than any type 
of Born or distorted wave Born approximation. The solution to the 
Lippmann--Schwinger equation is most conveniently obtained in a basis
independent formal language
\bea
\overline{G}_0(\omega)&=&\frac{1}{\pi\omega}\,{\cal P}\,
\int_0^\infty q^2 dq\,\left[
\frac{|\overline{\eta}_{\omega_q}\rangle\langle \overline{\eta}_{\omega_q}|}
{\omega-\omega_q}+
\frac{|\overline{\eta}_{-\omega_q}\rangle\langle \overline{\eta}_{-\omega_q}|}
{\omega+\omega_q}\right]\cr\cr
V(\omega)&=&-\frac{\omega_0}{\omega_\Theta-\omega}
|g_\Theta\rangle\langle g_\Theta|
\qquad {\rm with}\qquad
\langle r|g_\Theta\rangle=X_\Theta\left(
2\lambda(r)-\omega_0 M_K(r)\right)z(r)\,,
\label{lseq1}
\eea
where $\omega_p=\sqrt{p^2+m_K^2}$ and ${\cal P}$ denotes the principle 
value prescription. 
In the above expression $\overline{\eta}$
appears, rather than $\widetilde{\eta}$, because the interaction Hamiltonian
is treated as an additional interaction for the constrained fluctuations
and we thus require the Green's function for $\overline{\eta}$.
The potential is obtained from eq.~(\ref{potential})
by setting $X_\Lambda=0$. We find the reaction matrix
\be
R=\frac{\omega_0 X_\Theta^2|\left(2\lambda-\omega_0 M_K\right)z\rangle
\langle z\left(2\lambda-\omega_0 M_K\right)|}
{\omega-\omega_\Theta-\omega_0\langle z(2\lambda)|
\overline{G}_0(\omega)|(2\lambda)z\rangle}
\label{lsqe2}
\ee
since $\langle\overline{\eta}_\omega|M_K z\rangle=0$. The resonance
phase shift induced by $V$ is obtained from the diagonal matrix 
element of the R--matrix,
$k\langle\overline{\eta}_{\omega_k}|R|\overline{\eta}_{\omega_k}\rangle=
-{\rm tan}\left(\delta_{\rm R}(k)\right)$, yielding
\be
{\rm tan}\left(\delta_{\rm R}(k)\right)=
\frac{\Gamma(\omega_k)/2}{\omega_\Theta-\omega_k+\Delta(\omega_k)}\,.
\label{resformula}
\ee
This phase shift exhibits the canonical resonance structure with the
width
\be
\Gamma(\omega_k)=2k\omega_0 X_\Theta^2 
\left|\int_0^\infty r^2dr\, z(r)2\lambda(r)
\overline{\eta}_{\omega_k}(r)\right|^2
\label{width1}
\ee
and the pole shift
\be
\Delta(\omega_k)=\frac{1}{2\pi\omega_k}\,{\cal P}\,
\int_0^\infty q dq\,\left[
\frac{\Gamma(\omega_q)}{\omega_k-\omega_q}
+\frac{\Gamma(-\omega_q)}{\omega_k+\omega_q}\right]\,.
\label{poleshift}
\ee
Of course, we have numerically verified that in the large--$N_C$
limit with $X_\Theta^2=1$, the phase shift from 
eq.~(\ref{resformula}) is identical to the one that we calculated as 
the difference in eq.~(\ref{resps}). For finite $N_C$ the width 
$\Gamma$ acquires a factor $X_\Theta^2\ne1$ and the R--matrix formalism 
becomes two--dimensional~($\Lambda$ and $\Theta^+$ exchange).

We stress that although the singularity of the scattering amplitude 
is shifted by the amount $\Delta$ due to iterated resonance exchange, 
the "bare" mass of the resonance is $\omega_\Theta$, the pentaquark 
mass predicted in the RRA.

Alternatively we have applied Fermi's golden rule to compute the 
transition rate $\Theta\to KN$ from the interaction Hamiltonian, 
eq.~(\ref{hint}). This leads identically to the result displayed in 
eq.~(\ref{width1}) and shows that the way we have included the separable 
potential in eq.~(\ref{Yukawa1}) is unique.

\begin{figure}
\centerline{
\epsfig{file=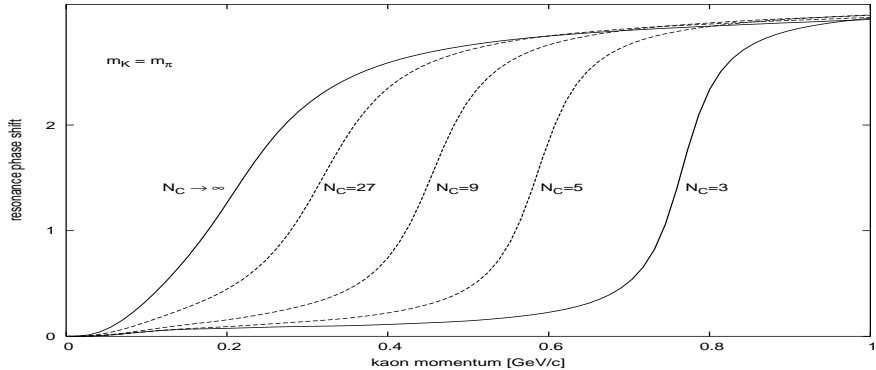,width=5cm,height=12cm,angle=270}}
\caption{\label{fig_5}The resonance phase shift as a function of $N_C$
for $m_K=m_\pi$.}
\end{figure}

\subsection{Finite $N_C$}

As already mentioned, the R--matrix formalism is two dimensional 
at finite $N_C$ because the $\Lambda$ exchange cannot be omitted.
It is therefore more convenient to compute the resonance phase shift in 
a fashion similar to eq.~(\ref{resps}). First we compute the phase
shift with both Yukawa interactions included from 
$\widetilde{\eta}$ in eq.~(\ref{Yukawa1}). From that we get the resonance 
phase shift by subtracting the background phase shift that we extract from 
$\overline{\eta}$ in eq.~(\ref{intdiff}).
The resulting resonance phase shifts in the $\Theta^+$ channel 
are displayed in figure~\ref{fig_5}. It is illuminating to see how
the BSA result emerges in the limit $N_C\to\infty$. Obviously this 
resonance becomes sharper and the pole shift decreases (in magnitude) to
$\Delta=-14{\rm MeV}$ as the number of color assumes its physical
value $N_C=3$. This is mainly due to the reduction of $X_\Theta$.
Furthermore, at finite $N_C$ the bare resonance position is increased
to $\omega_\Theta=(N_C+3)/4\Theta_K$, according to eq.~(\ref{Ntheta}).
This $N_C$--dependence of the resonance position is clearly 
exhibited in figure~\ref{fig_5}.

There are two competing effects on the width when approaching 
the physical value, $N_C=3$. 
The position of the pole is increased by about a factor of two.
This increases the available phase space and thus the width. On the
other hand, the collective coordinate matrix element $X_\Theta^2$ 
decreases by approximately a factor 3, {\it cf.\@} eq.~(\ref{xddr}). 
According to eq.~(\ref{width1}) this decreases the width.

At this point it is also illuminating to consider the large--$N_C$
expansion of these matrix elements in more detail. For example
\be
X_\Theta^2 \,\longrightarrow\, 1-\frac{7}{N_C}+\frac{52}{N_C^2}
-\frac{372}{N_C^3}+\frac{2628}{N_C^4}+\ldots\,\,,
\label{XlargeN}
\ee
does not converge for $N_C\le 7$.  This clearly demonstrates that
taking only leading pieces from a large--$N_C$ expansion of these
matrix elements yields incorrect results for $N_C=3$. It is unavoidable 
to include finite $N_C$ effects. Any approach~\cite{Je04} 
that employs $1/N_C$ expansion methods for exotic baryon matrix 
elements at $N_C=3$ seems questionable. 

\subsection{Comparison with other approaches to compute the width}

We stress that the appearance of only a single rather than 
two $SU(3)$ structures in the transition Hamiltonian in the flavor 
symmetric limit is not an artifact in the (eventually oversimplified)
Skyrme model whose action is a functional of only pseudoscalar 
mesons. Rather it generalizes to \emph{all} chiral soliton models. The
reason is simple. Starting point in all these models is the 
introduction of intrinsic fluctuations about the classical soliton,
{\it i.e.\@} the generalization of the {\it ansatz}, eq.~(\ref{Usu3})
to parameterize additional fields. In this intrinsic formulation 
the interaction Hamiltonian must be an isoscalar operator that 
is linear in kaonic fluctuations and projects onto the P--wave
channel. At ${\cal O}(N_C^0)$ only 
$d_{i\alpha\beta}\hat{x}_i\widetilde{\eta}_\alpha\Omega_\beta$ 
is possible. Of course, in more extended models $\widetilde{\eta}_\alpha$ 
also involves other strange degrees of freedom, {\it e.g.\@} $K^*$.
In the flavor symmetric formulation the adjoint representation of the
collective coordinates, $D_{ab}$, emerges \emph{only} by the 
transformation to the laboratory system as in eq.~(\ref{labframe}). To 
see whether or not the limitation to only a single $SU(3)$ structure
in the transition Hamiltonian is an artifact of the Skyrme model it is
instructive to recall how the authors of refs.~\cite{Di97,El04}
came to the improper use of several $SU(3)$ structures regardless of 
flavor symmetry breaking. Starting point of those studies is the
spatial part of the axial current and its identification with 
the kaon field operator via PCAC. This axial current is commonly
computed from the rigidly rotating classical field only and in general
allows for three $SU(3)$ structures in the flavor symmetric case.
One of those structures is directly related to the presumably 
small singlet current matrix element and is usually ignored in the
context of the $\Theta^+$ width. In the simple Skyrme 
model it vanishes exactly. However, the other two structures 
are also present in the Skyrme model axial current~\cite{Ka87}.
They simply do not show up in the transition Hamiltonian.
Stated otherwise, the axial current operator is not suitable to
compute hadronic decay widths in chiral soliton models. A simple
reason is that the isospin embedding of the hedgehog soliton, 
eq.~(\ref{hedgehog}) breaks flavor symmetry. In particular this 
prevents using $SU(3)$ to relate various hadron decay widths among 
each other. For example, the computation of the $\Delta\to\pi N$
width requires a term linear in the intrinsic pion fluctuations which
involves collective coordinate operators that are not related 
to~$d_{i\alpha\beta}\hat{x}_i\widetilde{\eta}_\alpha\Omega_\beta$.

Of course, in our approach 
$\partial^\mu A_\mu^a = f_\pi m_a^2 \phi^a + {\cal O}(\phi^3)$,
{\it i.e.\@} PCAC, holds as well. It is nothing but the equation of motion 
for the chiral angle and the meson fluctuations that we solve. 
That is, in our approach PCAC is satisfied at ${\cal O}(N_C)$
and ${\cal O}(N_C^0)$ while in refs.~\cite{Di97,El04} the 
equation of motion is only considered at ${\cal O}(N_C)$. The main
point, however, is that within soliton models the right hand side may 
not be the interpolating field for the pseudoscalar meson in the final 
state of hadronic decays.

\section{SU(3) symmetry breaking}

In section V we have established the equivalence of the bound state 
and rotation--vibration coupling approaches for computing the kaon 
nucleon phase shift in the large--$N_C$ limit in the flavor symmetric 
case. Here we will extend that discussion to the symmetry breaking 
case.  

The interaction Hamiltonian that describes the Yukawa
exchange for fluctuations in the laboratory frame,
{\it cf.} eq.~(\ref{hint}), acquires an additional term
from the symmetry breaking part of the action,~$\Gamma_{\rm SB}$,
\bea
H_{\rm int}&=&\frac{2f_\pi}{\Theta_K}
d_{i\alpha\beta}D_{\gamma\alpha}R_\beta
\int d^3r\,{\rm sin}\frac{F(r)}{2}
\left[2\lambda(r)-\omega_0 M_K(r)\right]
\widetilde{\xi}_{\gamma}(\vec{x\,},t)\hat{x}_i
\cr\cr &&\hspace{0.5cm}
-\frac{4f_\pi}{\sqrt3}
d_{i\alpha\beta}D_{\gamma\alpha}D_{8\beta}
\int d^3r\,{\rm sin}\frac{F(r)}{2}
\left[m_K^2-m_\pi^2-\frac{3\Gamma}{8\Theta_K}M_K(r)\right]
\widetilde{\xi}_{\gamma}(\vec{x\,},t)\hat{x}_i\,.
\label{hintsb}
\eea
We provide the detailed derivation of this interaction Hamiltonian in the 
appendix. In principle, there are many pole contributions, such
as from states with $S=\pm1$ and $I=0$ that stem from higher dimensional 
$SU(3)$ representations. They lead to sharp resonances at higher
energies and may be omitted for the energy regime of current interest.
We thus keep only the contributions associated with the exchanges of 
$\Lambda$ and $\Theta^+$. Then the separable Yukawa exchange potential 
in the P--wave has the general form
\bea
V(\omega)&=&-\frac{\omega_0}{\omega_\Theta-\omega}
|g_\Theta \rangle \langle g_\Theta|
-\frac{\omega_0}{\omega_\Lambda+\omega}
|g_\Lambda\rangle \langle g_\Lambda|
\cr\cr
\langle r|g_\nu\rangle&=&
X_\nu\left(2\lambda(r)-\omega_0M_K(r)\right)z(r)
+\frac{Y_\nu}{\omega_0}
\left(m_K^2-m_\pi^2-\frac{3\Gamma}{8\Theta_K}M_K(r)\right)z(r)\,,
\quad \nu=\Theta,\Lambda\,. \qquad
\label{vyukawa2}
\eea
Note that $\langle z | g_\nu\rangle=0$. This implies that the solutions
to the constrained equation of motion~(\ref{intdiff}) with this 
potential added, also do satisfy the constraint 
$\langle z |M_K\widetilde{\eta}\,\rangle=0$.

The matrix elements $X_\nu$ have already been defined in 
eq.~(\ref{ddr}). Similarly we introduce
\bea
\langle \Theta^+|d_{3\alpha\beta}D_{+\alpha}D_{8\beta}|n\rangle &=:&
Y_\Theta\sqrt{\frac{3}{8N_C}}\cr\cr
\langle \Lambda|d_{3\alpha\beta}D_{-\alpha}D_{8\beta}|p\rangle &=:&
Y_\Lambda\sqrt{\frac{3}{8N_C}}\,,
\label{ddd}
\eea
where $Y_\Theta$ and $Y_\Lambda$
tend to unity in the flavor symmetric case as $N_C\to\infty$.
Of course, for non--zero symmetry breaking the matrix elements 
$X_\nu$ and $Y_\nu$ have to be computed numerically 
and cannot be given as simple functions of $N_C$, results 
are listed in table~\ref{tab_1}. 

The integro--differential equation with the Yukawa pieces 
included is a bit more complicated than in the flavor 
symmetric case, eq.~(\ref{Yukawa1}),
\bea
h^2\widetilde{\eta}+\omega\left(2\lambda-\omega M_K\right)
\widetilde{\eta} &=& 
\cr\cr && \hspace{-3cm}
-\Big[2\lambda-\left(\omega+\omega_0\right)M_K
-\omega_0\left(\frac{X_\Theta^2}{\omega_\Theta-\omega}
+\frac{X_\Lambda^2}{\omega_\Lambda+\omega}\right)
\left(2\lambda-\omega_0M_K\right)
\cr\cr &&\hspace{-1.5cm}
-\left(\frac{X_\Theta Y_\Theta}{\omega_\Theta-\omega}
+\frac{X_\Lambda Y_\Lambda}{\omega_\Lambda+\omega}\right)
\left(m_K^2-m_\pi^2-\frac{3\Gamma}{8\Theta_K}M_K\right)\Big]z
\int_0^\infty dr r^2 z(2\lambda)\widetilde{\eta}
\cr\cr && \hspace{-3cm}
+\left(m_K^2-m_\pi^2\right)
\Big[M_K+\left(\frac{X_\Theta Y_\Theta}{\omega_\Theta-\omega}
+\frac{X_\Lambda  Y_\Lambda}{\omega_\Lambda+\omega}\right)
\left(2\lambda-\omega_0M_K\right)
\cr\cr && \hspace{-3cm}
+\frac{1}{\omega_0}
\left(\frac{Y_\Theta^2}{\omega_\Theta-\omega}
+\frac{Y_\Lambda^2}{\omega_\Lambda+\omega}\right)
\left(m_K^2-m_\pi^2-\frac{3\Gamma}{8\Theta_K}M_K\right)\Big]z
\int_0^\infty dr r^2 z\widetilde{\eta}\,.
\label{Yukawa1sb}
\eea
For simplicity we have omitted the arguments of the radial 
functions.

Let $\eta(r)$ again be a solution to the BSA equation~(\ref{scndorder}).
In the limit $N_C\to\infty$ we require that $\widetilde{\eta}=\eta-az$ with 
$a=\int_0^\infty dr r^2\, z(r) M_K(r)\eta(r)$ is a solution to 
eq.~(\ref{Yukawa1sb}) also in the case that symmetry breaking
($m_K\ne m_\pi$) is included. Modulo phase conventions 
this implies 
\be
X_\Theta=\frac{\omega_\Theta}{\omega_0} Y_\Theta\,, \qquad
X_\Lambda=-\frac{\omega_\Lambda}{\omega_0} Y_\Lambda\,, \qquad
Y_\Theta=Y_\Lambda=
\sqrt{\frac{\omega_0}{\omega_\Theta+\omega_\Lambda}}\,,
\label{XYcond}
\ee
as $N_C\to\infty$. And indeed, the matrix elements in 
eqs.~(\ref{ddr}) and~(\ref{ddd}) give these results when computed with 
the \emph{exact} solutions to eigenvalue equation~(\ref{evprob})
with flavor symmetry breaking included! 
In table~\ref{tab_1} we
display numerical results for these matrix elements for the physical
case $N_C=3$ and the large--$N_C$ limit for various values of $m_K$.
The third entry, $m_K=750{\rm MeV}$ yields the value of the symmetry
breaking parameter $\Gamma$ that we would have obtained if
we had included symmetry breaking terms that contain derivatives
of the chiral field.
Obviously, there are significant deviations
from the large--$N_C$ limit at $N_C=3$.
\begin{table}[h]
\centerline{\normalsize
\begin{tabular}{l c | c c | c c | c c}
 & $N_C$ & $X_\Theta$ & $Y_\Theta$ & $X_\Lambda$ & $Y_\Lambda$ 
& $X_{\Theta^*}$ & $Y_{\Theta^*}$ \cr
\hline
$m_K=m_\pi$ & $3$ &~~0.596~~&~~0.149~~&~~-0.377~~&~~0.189~~
&~~0.257~~&~~0.138~~\cr
&$\infty$ & 1 & 1 & 0 &1 & 0.667 & 0.667\cr
\hline
$m_K=495{\rm MeV}$ & 3 & 0.690 & 0.168 & -0.427 & 0.195 & 0.307 & 0.142\cr
&$\infty$ & 1.069 & 0.691 & -0.378 & 0.691 &0.713 & 0.461 \cr
\hline
$m_K=750{\rm MeV}$ & 3& 0.801 & 0.188 & -0.497 & 0.203 &0.376 & 0.147  \cr
&$\infty$ & 1.157 & 0.575 & -0.582 & 0.575 &0.771 & 0.383
\end{tabular}}
\caption{\label{tab_1}Matrix elements of collective coordinate
operators for $N_C=3$ and $N_C\to\infty$ and 
various values of $m_K$, {\it cf.\@} eqs.~(\ref{ddr}),~(\ref{ddd})
and~(\ref{XYstar0}). The third
entry, $m_K=750{\rm MeV}$ is chosen such as to compensate for the
omitted derivative type symmetry breaker.}
\end{table}

The large--$N_C$ equation of motion for $\widetilde{\eta}$ has a 
bound state solution in the anti--kaon channel 
exactly at the BSA energy $\omega_\Lambda=196{\rm MeV}$ (for 
$m_K=495{\rm MeV}$). This is a starting point to discuss corrections
to the RRA spectrum caused by the rotation--vibration coupling for
finite $N_C$ which improves on earlier attempts that introduced
fluctuations induced by the rigid rotations~\cite{We90}. 
However, that discussion is beyond the scope of this paper.

\section{Soliton model predictions for $\mathbf{\Theta^+}$}

\begin{figure}
\centerline{
\epsfig{file=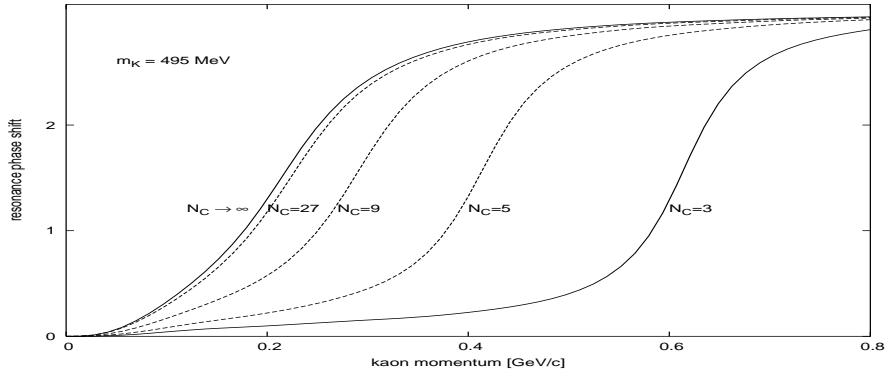,width=5cm,height=12cm,angle=270}}
\caption{\label{fig_6}The resonance phase shift as a function 
of $N_C$ for $m_K=495{\rm MeV}$.} 
\end{figure}

We have already noted that in the flavor symmetric case the 
excitation energy of $\Theta^+$ is altered according to 
eq.~(\ref{Ntheta}) for finite $N_C$.  The excitation energy
of $\Lambda$ remains at zero. For non--zero symmetry breaking
the situation is more complicated as both excitation energies
depend on the value for $m_K$. That is, we need to substitute
\be
\omega_\Lambda\, \longrightarrow\, E_\Lambda - E_N
\qquad {\rm and} \qquad
\omega_\Theta\, \longrightarrow\, E_\Theta - E_N\,,
\label{exeng}
\ee
where $E_{N,\Theta,\Lambda}$ are the solutions to the 
eigenvalue problem, eq.~(\ref{evprob}) in the respective 
channel for a prescribed $N_C<\infty$. In addition, we employ
the corresponding eigenstates of eq.~(\ref{evprob}) to 
evaluate the matrix elements in eqs.~(\ref{ddr}) and~(\ref{ddd})
to numerically compute $X_\Theta,X_\Lambda,Y_\Theta$ and $Y_\Lambda$
at any value of $N_C$ and $m_K$, {\it cf.\@} table~\ref{tab_1}. We 
then compute the resonance phase shift as described in section VI~B. 
The results are shown in figure~\ref{fig_6}.
Again these phase shifts exhibit a pronounced resonance structure.
As in the flavor symmetric case, finite $N_C$ effects cause the
pole position to move to higher momenta and also the resonance to 
become sharper. Although the phase shifts displayed in figures~\ref{fig_5}
and~\ref{fig_6} as a function of momentum seem to 
be similar this is not the case for the resonance energy since
it is subject to the dispersion relation $\omega=\sqrt{k^2+m_K^2}$
and $m_K$ differs significantly in the two case. From figure~\ref{fig_6}
we read off a $\Theta$--nucleon mass difference of about $750{\rm MeV}$
which is a bit higher than the empirical value of about $600{\rm MeV}$
if the $\Theta^+$ signal around $1540{\rm MeV}$ were indeed established.
We remark, however, that the actual prediction for this mass difference
is quite model dependent, {\it e.g.\@} we have omitted effects due
to $f_K\ne f_\pi$. Also, the inclusion of other fields like scalar--
or vector mesons or chiral quarks will alter the quantitative results.

The width 
function, $\Gamma(\omega)$ is computed from generalizing eq.~(\ref{width1}) to 
also contain symmetry breaking effects,
\be
\Gamma(\omega_k)=2k\omega_0
\left|X_\Theta\int_0^\infty r^2dr\, z(r)2\lambda(r)
\overline{\eta}_{\omega_k}(r)
+\frac{Y_\Theta}{\omega_0}\left(m_K^2-m_\pi^2\right)
\int_0^\infty r^2dr\,z(r)\overline{\eta}_{\omega_k}(r)\right|^2\,,
\label{widthsb}
\ee
which arise from the $\Theta^+$ matrix elements of the separable potential, 
eq.~(\ref{vyukawa2}).  Again, $\overline{\eta}$
rather than $\widetilde{\eta}$ appears in the above expression.
Note that contributions of the
form $\int dr r^2 z(r) M_K(r)\overline{\eta}(r)$ vanish due to the second
constraint in eq.~(\ref{constraints}).

\begin{figure}
\centerline{
\epsfig{file=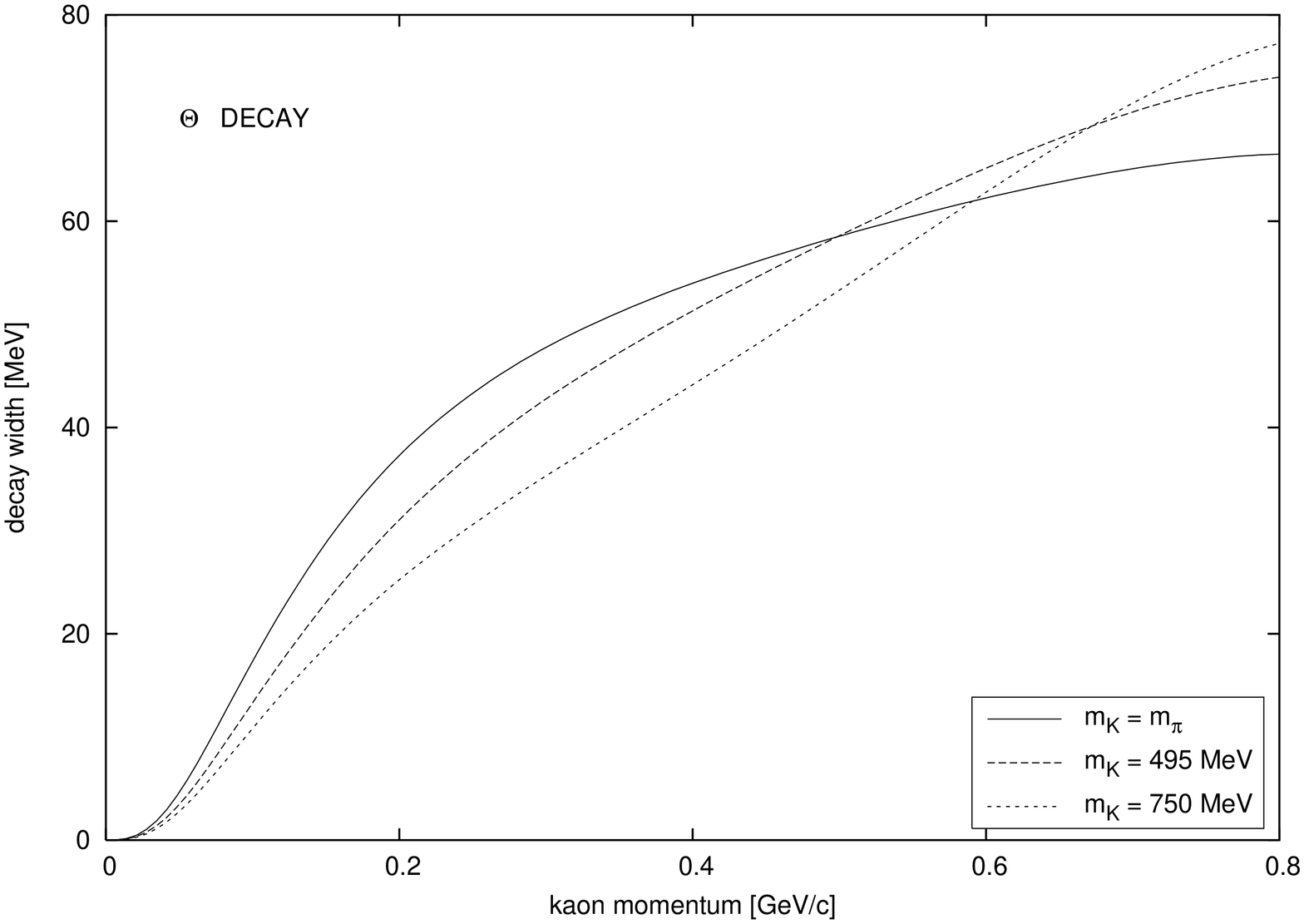,width=5cm,height=6cm,angle=270}\hspace{2cm}
\epsfig{file=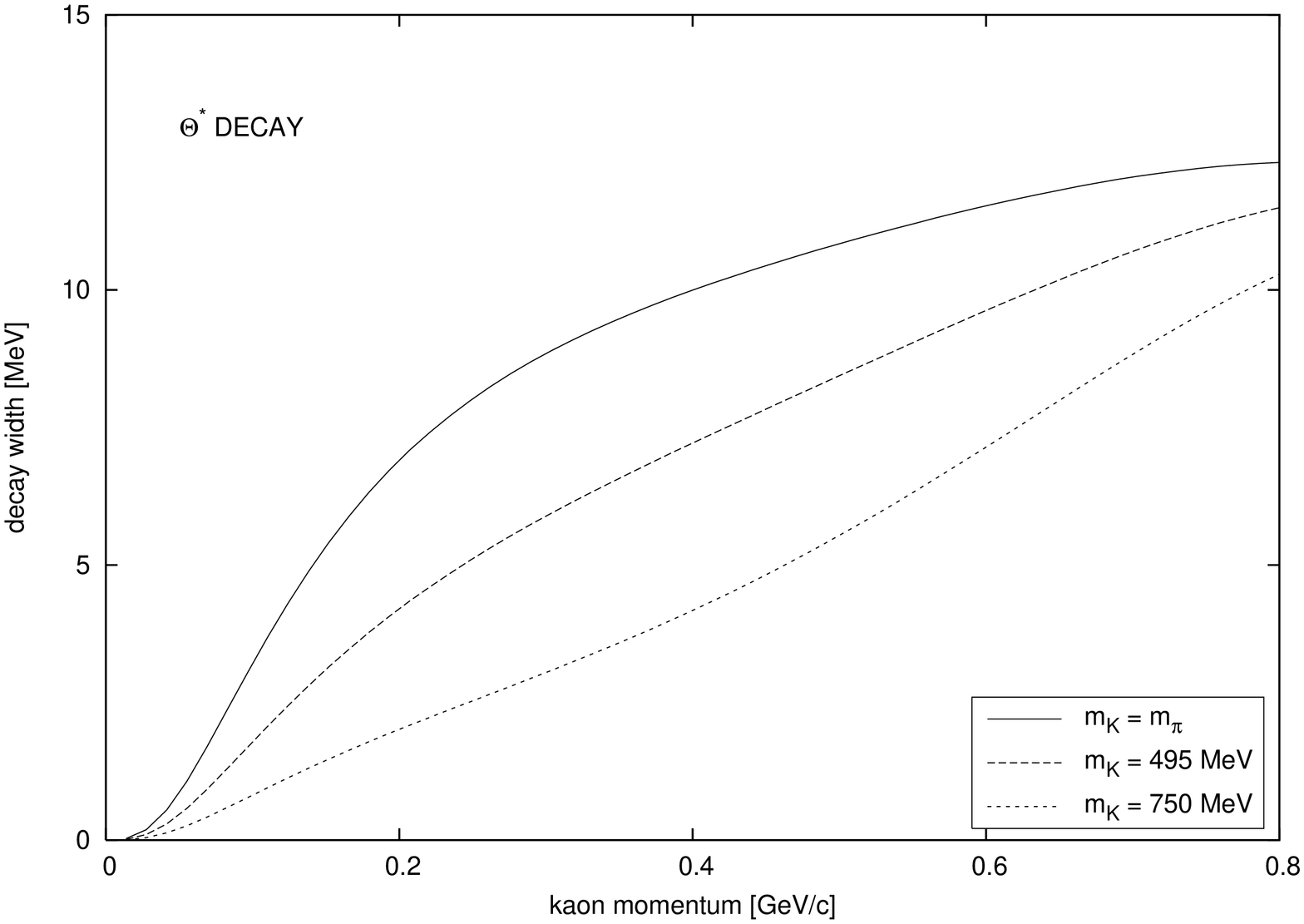,width=5cm,height=6cm,angle=270}}
\caption{\label{fig_7}Model prediction for the
width, $\Gamma(\omega)$ of $\Theta^+$ (left) and $\Theta^{*+}$ (right)
for $N_C=3$ as function of the momentum
$k=\sqrt{\omega^2-m_K^2}$ for three values of the kaon mass. 
Note that the two figures have different scales.}
\end{figure}

We have already mentioned that chiral soliton models only provide 
qualitative insight in the baryon mass spectrum. At best, we expect the 
model predictions for masses of exotic baryons to be reliable at the, say, 
$100{\rm MeV}$ level. Let us therefore now assume that the $\Theta^+$ 
resonance indeed corresponds to the recently asserted pentaquark 
of mass $M_\Theta=1540{\rm MeV}$ and estimate its width from $\Gamma(\omega)$.
First, we have to find the corresponding kaon momentum. This is not
without ambiguity because in the soliton picture baryon masses are 
of higher order in $N_C$ than the meson fluctuations and thus considered 
infinitely heavy. Therefore recoil effects are not necessarily included. 
This yields $k=340{\rm MeV}$ for the kaon momentum. Of course, a simple 
calculation using relativistic kinematics incorporates recoil effects 
and leads to $k=270{\rm MeV}$. For these two momenta we respectively read 
off widths of $47{\rm MeV}$ and $40{\rm MeV}$ for $m_K=495{\rm MeV}$ from 
figure~\ref{fig_7}. Fortunately the dependence of $\Gamma(\omega)$ on 
the effective strength of symmetry breaking seems only moderate in the 
resonance region. Hence the extracted width is not too sensitive on 
the uncertainties stemming from the omission of the derivative type 
symmetry breaker and we finally estimate $\Gamma\approx40{\rm MeV}$ as 
the width of the $\Theta^+$ pentaquark. 
Though we do not expect chiral soliton model predictions
for the widths to be reliable at the $10{\rm MeV}$ level,
we note that this result should be considered a small number as it has to
be compared to widths of other hadronic decays of baryon 
resonances, {\it e.g.\@} $\Gamma(\Delta\to\pi N)\approx120{\rm MeV}$,
empirically.

Finally let us briefly comment on the width of the 
$\Theta^{*}$ pentaquark, 
the $J=\frac{3}{2}$ and $I=1$ partner of $\Theta^+$. In large--$N_C$ 
the starting point is again the intrinsic $T$--matrix of the BSA. However, 
for $\Theta^{*}$ the recoupling between intrinsic laboratory frames
is more complicated than eq.~(\ref{pwaveint}). There are two 
intrinsic $T$--matrix elements $T_{{\cal G}\ell}=T_{\frac{1}{2}1}$ 
and $T_{\frac{3}{2}1}$ with different grand spins which contribute to 
the scattering in the laboratory frame~\cite{Ha84,Ma88},
\be
T(P13)=\frac{4}{9}\,T_{\fract{1}{2},1}
+\frac{5}{9}\,T_{\fract{3}{2},1}\,.
\label{recc}
\ee
While the collective excitation and thus the relevant phase shift
displayed in figures~\ref{fig_1} and~\ref{fig_2} emerges in the
intrinsic ${\cal G}=\frac{1}{2}$ channel, the ${\cal G}=\frac{3}{2}$
contributes only to the background phase shift and is irrelevant
for the $\Theta^*$ resonance structure. From eq.~(\ref{recc}) we notice
a factor $\frac{4}{9}$ not present in the corresponding relation
for $\Theta^+$, eq.~(\ref{pwaveint}). 

In the notation of the $RRA$ and for $m_K=m_\pi$ the $\Theta^*$ 
dwells in the $"\mathbf{27}"$--plet of flavor $SU(3)$ for finite $N_C$
while for $N_C\to\infty$ it becomes degenerate with $\Theta^+$. For 
$N_C=3$ the $\Theta^*$ pentaquark is expected to be roughly 
$100{\rm MeV}$ heavier than $\Theta^+$~\cite{Wa03,We04}. 
The Yukawa interactions
of the RVA, eq.~(\ref{hintsb}) now lead to (iterated) $\Theta^*$ and 
$\Sigma^*$ exchanges. 

The matrix elements relevant to compute the width of $\Theta^*$ are 
reduced via
\bea
\sum_{i=\pm\frac{1}{2}}
\langle \Theta^* |a^{(i)}_{1,m}d_{m\alpha\beta}
D_{\gamma(i)\alpha}R_\beta|
(KN)_{I=1,J=\frac{3}{2}}\rangle&=&\sqrt{3}\,
\langle \Theta^{*+} \uparrow| d_{3\alpha\beta}
D_{+\alpha}R_\beta|n \uparrow \rangle \cr \cr
\sum_{i=\pm\frac{1}{2}}
\langle \Theta^* |a^{(i)}_{1,m}d_{m\alpha\beta}
D_{\gamma(i)\alpha}D_{8\beta}|
(KN)_{I=1,J=\frac{3}{2}}\rangle&=&\sqrt{3}\,
\langle \Theta^{*+} \uparrow| d_{3\alpha\beta}
D_{+\alpha}D_{8\beta}|n \uparrow \rangle\,.
\label{WE2}
\eea
The relative factor $1/\sqrt{2}$ in comparison with eq.~(\ref{WE1}) is
essential so that for $\Theta^{*}$ we have to replace 
$X_\Theta\to X_{\Theta^*}$ and $Y_\Theta\to Y_{\Theta^*}$ in 
eq.~(\ref{widthsb}) by
\be
X_{\Theta^*}=\frac{4}{\sqrt{N_C}}\,
\langle \Theta^{*+} \uparrow| d_{3\alpha\beta}
D_{+\alpha}R_\beta|n \uparrow \rangle
\quad {\rm and}\quad
Y_{\Theta^*}=2\sqrt{\frac{N_C}{3}}\,
\langle \Theta^{*+} \uparrow| d_{3\alpha\beta}
D_{+\alpha}D_{8\beta}|n \uparrow \rangle\,.
\label{XYstar}
\ee
In the flavor symmetric case we find 
\be
X_{\Theta^*}=\frac{2}{3}\left(N_C-1\right)
\sqrt{\frac{N_C+5}{N_C(N_C+3)(N_C+9)}}=\,\frac{2}{3}\,
\begin{cases}
\frac{2}{9}\sqrt{3} &, \quad N_C=3\cr
\,\,1 &, \quad N_C=\infty\cr
\end{cases}\,.
\label{XYstar0}
\ee
Thus, disregarding phase space effects (due to $\Theta^{*}$ 
lying roughly $100{\rm MeV}$ above $\Theta^+$ for $N_C=3$), we
obtain
\be
\Gamma_{\Theta^*}=\Gamma_\Theta\left(\frac{X_{\Theta^*}}{X_\Theta}\right)^2
=\Gamma_\Theta
\begin{cases}
\,\frac{5}{27} &, \quad N_C=3\cr
\,\frac{4}{9} &, \quad N_C=\infty\cr
\end{cases}\,.
\label{widthstar}
\ee
From eq.~(\ref{XYstar0}) we notice that in the large--$N_C$ limit the 
factor $\frac{4}{9}$ that appears in eq.~(\ref{recc}) is recovered. This 
is precisely what is required in order to obtain the BSA result in this
limit. We have numerically verified that in the symmetry breaking
case
\be
\lim_{N_C\to\infty} \frac{X_{\Theta^*}}{X_\Theta}
=\lim_{N_C\to\infty} \frac{Y_{\Theta^*}}{Y_\Theta}
=\frac{2}{3}
\label{XYstar1}
\ee
for any value $m_K\ge m_\pi$ such that 
$\Gamma_{\Theta^{*}}=\frac{4}{9}\Gamma_\Theta$ in that case as well,
see also table~\ref{tab_1}.
Furthermore from eq.~(\ref{widthstar}) we also notice that for 
$N_C=3$ the $\Theta^{*}$ width is even more suppressed in comparison
to $\Theta^+$ (by roughly a factor of five in the symmetric case.)
Finite symmetry breaking lowers the prediction for the $\Theta^{*}$ width
even further, {\it cf.\@} figure~\ref{fig_7}. Thus we expect
a very sharp $\Theta^{*}$ resonance with a width of roughly $10{\rm MeV}$
about $100{\rm MeV}$ above the $\Theta^+$ pentaquark. 
This is opposite to the above criticized scenario based on matrix elements 
of the axial current to compute widths for hadronic decays, {\it e.g.\@}
the authors of ref.~\cite{El04} obtained 
$\Gamma_{\Theta^{*}}/\Gamma_{\Theta}=3...6$.

\section{Conclusions}

We have thoroughly compared the bound state (BSA) and the collective 
coordinate approaches to chiral soliton models in flavor $SU(3)$. For
definiteness we have only considered the simplest version of the 
Skyrme model augmented by the Wess--Zumino and symmetry breaking
terms. However, our analysis merely concerns the quantization
of fluctuations about the classical soliton that generates baryons 
states. Therefore our qualitative results are valid for \emph{any}
chiral soliton model.

Often the collective coordinate approach is identified with the 
rigid rotator approach (RRA) as only quantizing the spin--flavor
orientation of the soliton. But, a sensible comparison with
the BSA requires the consideration of harmonic oscillations in the
collective coordinate approach as well. This generalizes to the 
rotation--vibration approach (RVA). Here we have studied that 
scenario exhaustively and in particular ensured that the introduction 
of such fluctuations does not double--count any degrees of freedom. 
Only in this way the model generates an interaction Hamiltonian that 
describes hadronic decays. In doing so, we have solved the long
standing Yukawa problem in the kaon sector.
Technically the derivation of this Hamiltonian
is quite involved, however, the result is as simple as convincing:
In the limit $N_C\to\infty$, in which the BSA is undoubtedly correct,
the RVA and BSA yield identical results for both the baryon 
spectrum as well as the kaon--nucleon $S$-matrix. This equivalence
also holds when flavor symmetry breaking is included. In the 
first place this ensures that we have correctly introduced the 
collective coordinates. On top of that, this result
is very encouraging as it clearly demonstrates that collective
coordinate quantization is valid regardless of whether or not
the respective modes are zero--modes. Though the large--$N_C$ limit
is helpful for testing the results of the RVA, we have also 
seen that taking only leading terms in the respective matrix 
elements is not trustworthy.

There has been quite some confusion from the potential
disagreements between the BSA and the RVA. There are essentially
two reasons for that confusion:
\begin{itemize}
\item[1)] The BSA yields a slowly rising phase shift in the 
$S=+1$ channel that led to the misinterpretation that no resonance
was present. However, the scenario is that there is indeed a (broad) 
resonance which, unfortunately, is hidden by a repulsive background.
\item[2)] The $\Theta^+$ width was incorrectly computed.
Those computations did not even attempt to make contact
with the BSA.
\end{itemize}
Here we have resolved these puzzles: We observe a relatively
sharp resonance in the RVA for $N_C=3$ that evolves into the broad
structure seen in the BSA as $N_C$ increases.

Another major result of our calculation is that in the flavor symmetric
case the interaction Hamiltonian contains only a single leading $N_C$
structure of $SU(3)$ matrix elements for the $\Theta^+\to KN$ transition.
Any additional structure only enters via flavor symmetry breaking
terms in the Lagrangian. This is at odds with earlier approaches 
that essentially assumed any possible structure and fitted coefficients 
from a variety of hadronic decays under the assumption of $SU(3)$ 
relations. We stress that this is not a valid procedure as already 
the embedding of the classical soliton breaks $SU(3)$ and thus yields 
different structures for different hadronic transitions. In particular 
strangeness conserving and strangeness changing processes are not related 
to each other in chiral soliton model treatments. As for $m_K=m_\pi$
only a single transition operator exists, there cannot occur any cancellation 
between different (leading $N_C$) structures to explain the probably
small $\Theta^+$ width.

Though these general results are model independent, the numerical
results that we obtain for the masses and the widths of pentaquarks
are not. Moreover, the prediction for the latter may suffer from those
subleading $1/N_C$ contributions that were not considered here. Our 
numerical estimates indicate that the $\Theta^+$ is roughly $700{\rm MeV}$ 
heavier than the nucleon with quite a small width of about 
$40{\rm MeV}$. We predict the $\Theta^*$ to be about $100{\rm MeV}$
heavier than the $\Theta^+$ with an even smaller width of the order 
of $10{\rm MeV}$. It should be kept in mind, that all quantitative 
results for masses and widths are model dependent and that their
accurate predictions are very delicate. Despite of that, chiral soliton
models do predict the existence of the low--lying pentaquarks
$\Theta$ and $\Theta^*$ with strangeness, $S=+1$ and isospin, $I=0,1$. 
This is irrespective of details of the model and/or the adopted approach.

Let us finally speculate about implications that indications for 
pentaquarks seen earlier in kaon--nucleon scattering~\cite{PDG86,Hy92} 
may have on chiral soliton models.  At that time the structures observed 
in the $S=+1$ channel were assigned the quantum numbers $P_{01}(1830)$, 
$P_{13}(1810)$, $D_{03}(1790)$ and $D_{15}(2070)$.  In chiral soliton 
models the latter can only be interpreted as a quadrupole excitation of 
a rotational ground state in the $P_{01}$ channel. The structure observed
in that channel is actually higher in energy than both the $P_{13}$ and
$D_{03}$ structures. It is thus suggestive that this $P_{01}$ structure
should not be identified with the rotational ground state whose mass 
should thus be significantly lower than $1790{\rm MeV}$. This would be 
a further hint for a $\Theta^+$ pentaquark with 
$M_\Theta\lesssim 1700{\rm MeV}$.

\section*{Acknowledgment}
We very much appreciate interesting and helpful discussions
with G. Holzwarth and  V. Kopeliovich.

\section*{Note added}
After this manuscript had been submitted, it has been commented
on by T. Cohen~\cite{Co05}. We have rebutted his arguments in 
a reply~\cite{Wa05}.

\appendix
\section{Constrained Equations of Motion}

In this appendix we will derive the equation of motion~(\ref{intdiff})
for the fluctuations in the RVA. We find it more convenient to perform
that analysis in a general language for $\widetilde{\eta}_\alpha$ rather 
than for the P--wave 
projection, eq.~(\ref{pwave}). In that language the kaonic zero mode 
wave--function carries two indices. They parameterize the rotation
of kaon degrees of freedom ($\alpha=4,\ldots,7$) into any direction
($a=1,\ldots,8$) of $SU(3)$,
\be
z_\alpha^a(\vec{x\,})=\frac{2f_\pi}{\sqrt{\Theta_K}}\,
{\rm sin}\frac{F(r)}{2}\,\hat{x}_i f_{i\alpha a}\,.
\label{su3zm}
\ee
Only four of these wave--functions ($b=\beta=4,\ldots,7$) are 
non--zero. They correspond to different choices of the 
zero--mode analogue of the complex isospinor in eq.~(\ref{pwave})
and are normalized such that
$\int d^3r \,z_\alpha^{\beta}(\vec{x\,})M_K(r)
z_\alpha^{\gamma}(\vec{x\,})=\delta_{\beta\gamma}$
where $M_K(r)$ is the metric function defined in eq.~(\ref{radfct}). 
The analogue of eq.~(\ref{defom0}) reads
\be
\int d^3r\,\, 2\lambda(r)\, f_{8\alpha\beta}
z_\alpha^{\gamma}(\vec{x\,})z_\beta^{\delta}(\vec{x\,})=
\omega_0\,f_{8\gamma\delta}\,.
\label{su3defom0}
\ee
We postpone the discussion of the rotation--vibration coupling
terms involving the angular velocity $\Omega_8$ to the end of 
this appendix and collect all other terms in the Lagrange function 
that are ${\cal O}(N_C^0)$ according to the rules, eq.~(\ref{ordOmega}).
The result of this tedious calculation is
\bea
L&=&-M-\frac{N_C}{2\sqrt 3}\,\Omega_8
+\frac{1}{2}\Theta_K\Omega_\alpha^2
+\frac{1}{2}\int d^3r\left[M_K(r)\,
\dot{\widetilde{\eta}}_\alpha(\vec{x\,})\,
\dot{\widetilde{\eta}}_\alpha(\vec{x\,})
-\widetilde{\eta}_\alpha(\vec{x\,})\,
h^2_{\alpha\beta}(\vec{x\,})\,
\widetilde{\eta}_\beta(\vec{x\,})\right]
\cr\cr && \hspace{1.5cm}
+\frac{2}{\sqrt{3}}\int d^3r \lambda(r)\, f_{8\alpha\beta}\,
\widetilde{\eta}_\alpha(\vec{x\,})\,
\dot{\widetilde{\eta}}_\beta(\vec{x\,})
\cr\cr && \hspace{1.5cm}
-\sqrt{\Theta_K}\,\Omega_\alpha \int d^3r \,
z^\alpha_\beta(\vec{x\,})\left[M_K(r)\,
\dot{\widetilde{\eta}}_\beta(\vec{x\,})
-\frac{4}{\sqrt{3}}\lambda(r)\,f_{8\beta\gamma}\,
\widetilde{\eta}_\gamma(\vec{x\,})\right]
\cr\cr && \hspace{1.5cm}
+\frac{4}{3}\sqrt{\Theta_K}
\left(m_K^2-m_\pi^2\right) f_{8\alpha\beta}D_{8\alpha}
\int d^3r\, z_\gamma^\beta(\vec{x\,})\,
\widetilde{\eta}_\gamma(\vec{x\,})
+{\cal O}\left(\frac{1}{\sqrt{N_C}}\right)\,,
\label{Lorder0}
\eea
where dots denote derivatives with respect to time.
Since we are mainly interested in relations between conjugate momenta, 
the explicit form of the hermitian potential $h^2_{\alpha\beta}(\vec{x\,})$
(that also contains the spatial derivative operators as indicated in 
eq.~(\ref{radfct})) is irrelevant. For the Skyrme model Lagrangian, 
eq.~(\ref{lag}) it may be traced from the literature, {\it e.g.\@} 
refs.~\cite{Ca85,Scha94}.
The last two terms in eq.~(\ref{Lorder0}) are new and describe the
coupling between fluctuations and the collective coordinates.
The very last term would be absent in a flavor symmetric world. In that
case only a single $SU(3)$ operator, that is linear in the kaonic
angular velocity $\Omega_\alpha$, couples the fluctuations to the 
collective coordinates. We will soon see that in the contribution
proportional to $\Omega_\alpha$ in eq.~(\ref{Lorder0}) only the piece that
stems from the Wess--Zumino term survives because of the constraints.

The conjugate momenta of the kaon degrees of freedom are
\bea
R_\alpha&=&-\frac{\partial L}{\partial \Omega_\alpha}
=-\Theta_K\,\Omega_\alpha 
+\sqrt{\Theta_K}\int d^3r\, z_\beta^\alpha(\vec{x\,})\left[M_K(r)
\dot{\widetilde{\eta}}_\beta(\vec{x\,})-\frac{4}{\sqrt{3}} \lambda(r) 
f_{8\beta\gamma}\widetilde{\eta}_{\gamma}(\vec{x\,})\right]\,,
\\ \label{Rconjmom}
\Pi_\alpha(\vec{x\,})&=&
\frac{\delta L}{\delta \dot{\widetilde{\eta}}_\alpha(\vec{x\,})}
=-\sqrt{\Theta_K}\,\Omega_\beta M_K(r) z_\alpha^\beta(\vec{x\,})
+M_K(r)\dot{\widetilde{\eta}}_\alpha(\vec{x\,})
-\frac{2}{\sqrt{3}}\lambda(r)
f_{8\beta\gamma}\widetilde{\eta}_\gamma(\vec{x\,})\,.
\label{piconjmom}
\eea
Obviously, they are linearly dependent,
\be
R_\alpha=\sqrt{\Theta_K}\int d^3 r\, z_\beta^\alpha(\vec{x\,})\left[
\Pi_\beta(\vec{x\,})-\frac{2}{\sqrt{3}}
\lambda(r)f_{8\beta\gamma}\widetilde{\eta}_\gamma(\vec{x\,})\right]\,.
\label{lineardep}
\ee
The second piece stems from the Wess--Zumino term. 
It is unexpected because the conjugate momenta 
originate from parts in the Lagrangian that contain time derivatives
of the chiral field,~$\dot{U}$. In turn the linear dependence between the 
conjugate momenta results from the connection between
$\frac{\partial\dot U(\vec{x\,},t)}{\partial \Omega_\alpha}$ and
$\frac{\delta\dot U(\vec{x\,},t)}
{\delta \dot{\widetilde{\eta}}_\alpha(\vec{x\,})}$
causing a linear dependence between $R_\alpha$ and $\Pi_\alpha$ only.
However this argument only holds for \emph{local} pieces of the 
action while the Wess--Zumino term is \emph{non--local}.

We disentangle the collective and fluctuation pieces from the 
$\overline{\eta}$ momenta
\be
\Pi_\alpha^{\rm (coll)}(\vec{x\,})=
-\sqrt{\Theta}_K\,\Omega_\beta M_K(r) z^\beta_\alpha(\vec{x\,})
\quad {\rm and} \quad
\widetilde{\Pi}_\alpha(\vec{x\,})=
M_K(r)\dot{\widetilde{\eta}}_\alpha(\vec{x\,})-\frac{2}{\sqrt{3}}
\lambda(r)f_{8\alpha\beta}\widetilde{\eta}_\beta(\vec{x\,})
\label{collpart}
\ee
and demand that the momenta conjugate to the collective coordinates
do not contain any fluctuation parts,
\be
R_\alpha=\sqrt{\Theta}_K\,\int d^3r\, z^\alpha_\beta(\vec{x\,})
\Pi^{\rm (coll)}_\beta(\vec{x\,})=-\Theta_K\Omega_\alpha\,.
\label{flucfree}
\ee
The linear relation, eq.~(\ref{lineardep}) then translates into the 
\emph{primary constraints}
\be
\Psi_\alpha=\int d^3r\, z^\alpha_\beta(\vec{x\,}) \left[
\widetilde{\Pi}_\beta(\vec{x\,})-\frac{2}{\sqrt{3}}
\lambda(r)f_{8\beta\gamma}\widetilde{\eta}_\gamma(\vec{x\,})\right]=0\,.
\label{primary}
\ee
The corresponding \emph{secondary constraints} require the fluctuations to 
be orthogonal to the zero--mode
\be
\chi_\alpha=\int d^3r\, z^\alpha_\beta(\vec{x\,})\,
M_K(r)\widetilde{\eta}_\beta(\vec{x\,})=0\,.
\label{secondary}
\ee
These constraints are linear functionals of the fluctuations 
and their conjugate momenta. They satisfy the Poisson brackets
$\{\Psi_\alpha,\chi_\beta\}=\delta_{ab}$ so that $\widetilde{\Pi}_\alpha$ and
$\widetilde{\eta}_\beta$ are conjugate to each other in the 
constrained subspace~\cite{Di50}. This is unaltered by the 
(unexpected) contribution to $\Psi_\alpha$ from the Wess--Zumino term.
After projection onto the P--wave channel the constraints, 
eqs.~(\ref{primary}) and~(\ref{secondary}) turn into constraints for 
the associated radial functions given in eq.~(\ref{constraints}).

We need to find the equations of motion for $\widetilde{\Pi}_\alpha$
and $\widetilde{\eta}_\alpha$ and also to extract the Hamiltonian for
the interaction between the collective and fluctuation degrees of
freedom. We start by the Legendre transformation and add the
constraints $\Psi_\alpha$ and $\chi_\alpha$ with Lagrange multipliers 
$\alpha_\alpha$ and $\beta_\alpha$ respectively,
\be
H=-L-R_\alpha\Omega_\alpha 
+\int d^3r\, \Pi_\alpha(\vec{x\,})\,\dot{\widetilde{\eta}}_\alpha(\vec{x\,})
+\alpha_\alpha \Psi_\alpha + \beta_\alpha \chi_\alpha\,.
\label{legendre}
\ee
We express this ${\cal O}(N_C^0)$ Hamiltonian in terms of the 
conjugate fields $\widetilde{\Pi}_\alpha$ and $\widetilde{\eta}_\alpha$
as well as the angular momenta $R_\alpha$,
\bea
H&=&\frac{R_\alpha^2}{2\Theta_K}
+\frac{1}{2}\int \frac{d^3r}{M_K(r)} \, \left[
\widetilde{\Pi}_\alpha(\vec{x\,})+\frac{2}{\sqrt{3}}\lambda(r)
f_{8\alpha\beta}\,\widetilde{\eta}_\beta(\vec{x\,})\right]^2
+\frac{1}{2} \int d^3r\, \widetilde{\eta}_\alpha(\vec{x\,})
h^2_{\alpha\beta}(\vec{x\,})\widetilde{\eta}_\beta(\vec{x\,})\cr\cr
&&\hspace{0.5cm}
+\frac{R_\alpha}{\sqrt{\Theta_K}} \int d^3r\,
z_\beta^\alpha(\vec{x\,})\left[\widetilde{\Pi}_\alpha(\vec{x\,})
+\frac{2}{\sqrt{3}}\lambda(r)f_{8\alpha\beta}\,
\widetilde{\eta}_\beta(\vec{x\,})\right]
+\alpha_\alpha \Psi_\alpha \cr\cr
&&\hspace{0.5cm}
-\frac{4}{3}\sqrt{\Theta_K}
\left(m_K^2-m_\pi^2\right) f_{8\alpha\beta}D_{8\alpha}
\int d^3r\, z_\gamma^\beta(\vec{x\,})\,
\widetilde{\eta}_\gamma(\vec{x\,})
+\beta_\alpha \chi_\alpha\,.
\label{fullH}
\eea
The Hamiltonian obviously contains terms that are explicitly linear 
in the fluctuations or their conjugate momenta. These terms contribute
to the Yukawa interaction, eq.~(\ref{hintsb}). Additional linear terms 
arise from the Lagrange--multipliers which we compute in two steps 
from the equations of motion 
\bea
\dot{\widetilde{\eta}}_\alpha(\vec{x\,})&=&
\frac{\delta H}{\delta \widetilde{\Pi}_\alpha(\vec{x\,})}
=\frac{1}{M_K(r)}\left[\widetilde{\Pi}_\alpha(\vec{x\,})
+\frac{2}{\sqrt{3}}\lambda(r)f_{8\alpha\beta}\,
\widetilde{\eta}_\beta(\vec{x\,})\right]
+\left(\frac{R_\beta}{\sqrt{\Theta_K}}+\alpha_a\right)
z_\alpha^\beta(\vec{x\,})\cr\cr
\dot{\widetilde{\Pi}}_\alpha(\vec{x\,})&=&
\frac{\delta H}{\delta \widetilde{\eta}_\alpha(\vec{x\,})}=
-h^2_{\alpha\beta}(\vec{x\,})\widetilde{\eta}_\beta(\vec{x\,})
+\frac{2}{\sqrt{3}}\frac{\lambda(r)}{M_K(r)}
f_{8\alpha\beta}\left[
\widetilde{\Pi}_\alpha(\vec{x\,})+\frac{2}{\sqrt{3}}\lambda(r)
f_{8\alpha\beta}\,\widetilde{\eta}_\beta(\vec{x\,})\right] \cr\cr
&&\hspace{0.5cm}
+\frac{2}{\sqrt{3}}\left(\frac{R_\beta}{\sqrt{\Theta_K}}-\alpha_a\right)
\lambda(r)f_{8\alpha\gamma}z_\gamma^\beta(\vec{x\,})
+\frac{4}{3}\sqrt{\Theta_K} \left(m_K^2-m_\pi^2\right) 
f_{8\beta\gamma}D_{8\beta} z_\alpha^\gamma(\vec{x\,})\cr\cr
&&\hspace{0.5cm}
-\beta_\beta M_K(r)z_\alpha^\beta(\vec{x\,})\,.
\label{eqofmotion}
\eea
First, the constraints, eq.~(\ref{primary}) 
and~(\ref{secondary}) must be valid for all times. Therefore their 
derivatives with respect to time 
\bea
\dot{\Psi}_\alpha&=&\int d^3r\, z^\alpha_\beta(\vec{x\,}) \left[
\dot{\widetilde{\Pi}}_\beta(\vec{x\,})-\frac{2}{\sqrt{3}}
\lambda(r)f_{8\beta\gamma}\,\dot{\widetilde{\eta}}_\gamma(\vec{x\,})\right]=0
\cr\cr
\dot{\chi}_\alpha&=&\int d^3r\, z^\alpha_\beta(\vec{x\,})\,
M_K(r)\dot{\widetilde{\eta}}_\beta(\vec{x\,})=0
\label{constraintderiv}
\eea
must also vanish. Second, we substitute the equations of 
motion~(\ref{eqofmotion}) into these relations and extract the 
Lagrange--multipliers,
\bea
\alpha_\alpha&=&-\frac{R_\alpha}{\sqrt{\Theta_K}}
-\frac{4}{\sqrt{3}}\int d^3r\, \lambda(r) z_\beta^\alpha(\vec{x\,})
f_{8\beta\gamma}\widetilde{\eta}_\gamma(\vec{x\,})\cr\cr
\beta_\alpha&=&-\frac{2}{\sqrt{3}}\,\omega_0\,f_{8\alpha\beta}\alpha_\beta
-\frac{\Gamma}{2\sqrt{\Theta_K}}f_{8\alpha\beta}D_{8\beta}
-\left(m_K^2-m_\pi^2\right)\int d^3r\, 
z_\beta^\alpha(\vec{x\,})\widetilde{\eta}_\beta(\vec{x\,})\,,
\label{lagmultiapp}
\eea
with $\omega_0$ and $\Gamma$ defined in eqs.~(\ref{defom0}) 
and~(\ref{defGamma}), respectively. As the conjugate momenta,
these Lagrange--multipliers obviously carry collective 
\be
\alpha^{\rm (coll)}_\alpha=-\frac{R_\alpha}{\sqrt{\Theta_K}}
\quad {\rm and} \quad
\beta^{\rm (coll)}_\alpha=
\frac{2}{\sqrt{3}}\,\omega_0\,f_{8\alpha\beta}\frac{R_\beta}{\sqrt{\Theta_K}}
-\frac{\Gamma}{2\sqrt{\Theta_K}}f_{8\alpha\beta}D_{8\beta}
\label{colllagmulti}
\ee
as well as fluctuation pieces: 
$\widetilde{\alpha}_\alpha=\alpha_\alpha-\alpha^{\rm (coll)}_\alpha$
and $\widetilde{\beta}_\alpha=\beta_\alpha-\beta^{\rm (coll)}_\alpha$.
When substituted into the Hamiltonian, eq.~(\ref{fullH}) the 
collective pieces provide the above mentioned additional contribution
to the Yukawa interaction. In total the latter reads,
\bea
H_{\rm int}&=&\frac{2R_\alpha}{\sqrt{3\Theta_K}}
f_{8\beta\gamma}\int d^3r\, 
z_\beta^\alpha(\vec{x\,})\left[2\lambda(r)-\omega_0 M_K(r)\right]
\widetilde{\eta}_\gamma(\vec{x\,})\cr\cr
&&\hspace{0.2cm}
-\frac{4}{3}\sqrt{\Theta_K} D_{8\alpha}f_{8\alpha\beta}\int d^3r\,
z^\beta_\gamma(\vec{x\,})
\left[m_K^2-m_\pi^2-\frac{3\Gamma}{8\Theta_K}M_K(r)\right]
\widetilde{\eta}_\gamma(\vec{x\,})\,.
\label{hintapp}
\eea
The formula, eq.~(\ref{hintsb}) displayed in the main part of the paper 
finally results from replacing the zero--mode wave--function according to 
eq.~(\ref{su3zm}) and the intrinsic fluctuations according to 
eq.~(\ref{labframe1}). 

We abstain from presenting further details on the 
homogeneous parts of the equations of motion for $\widetilde{\Pi}_\alpha$
and $\widetilde{\eta}_\alpha$. It completely suffices to note that 
by (i) projecting on the P--wave channel, (ii) omitting the explicitly
inhomogeneous pieces proportional to $R_\alpha$ and $D_{8\alpha}$ (flavor 
symmetry breaking) and (iii) substituting the fluctuation pieces
for the Lagrange--multipliers: 
$\alpha_\alpha\to\widetilde{\alpha}_\alpha$ and
$\beta_\alpha\to\widetilde{\beta}_\beta$, eqs.~(\ref{eqofmotion})
directly transform into eqs.~(\ref{blaizoteqmconst}) that we 
employed to compute the background phase shift in section III.
In the main text the solutions of these homogeneous parts are denoted
$\overline{\Pi}_\alpha$ and $\overline{\eta}_\alpha$, respectively.

Finally we would like to comment on the treatment of the angular
velocity $\Omega_8$ and its $N_C$ scaling. To gain some insight, we 
compute the time derivative of $R_\alpha$ from $[R_\alpha,H]$ with $H$ 
given in eq.~(\ref{hcol}),
\be
\frac{d R_\alpha}{dt}=
\left(\frac{1}{\Theta_\pi}-\frac{1}{\Theta_K}\right)
f_{i\alpha\beta}R_iR_\beta
-\left(\frac{R_8}{\Theta_K}+\Omega_8\right)
f_{8\alpha\beta}R_\beta
-\frac{\Gamma}{2}f_{8\alpha\beta}D_{8\beta}\,.
\label{Radot}
\ee
While the first term, which is subject to ordering ambiguities, 
is suppressed by ${\cal O}(1/\sqrt{N_C})$, the second
term, together with the constraint $R_8=\frac{N_C}{2\sqrt{3}}$,
suggests that $\Omega_8$ should be counted as ${\cal O}\left(N_C^0\right)$
and thus not be omitted as $N_C\to\infty$. 

By isolating the $\Omega_8$ contribution from the time derivative 
of the {\it ansatz}, eq.~(\ref{Usu3}) we find to linear order
in the fluctuations,
\be
\frac{dU(\vec{x\,},t)}{dt}=\frac{i}{f_\pi}
\left(\frac{2}{\sqrt{3}}\,\omega+\Omega_8\right)
A_3(t)\sqrt{U_0(\vec{x\,})}\,
\left[f_{8\alpha\beta}\lambda_\alpha
\widetilde{\eta}_\beta(\vec{x\,},t)\right]
\sqrt{U_0(\vec{x\,})}\, A_3(t)^\dagger
+{\cal O}(\widetilde{\eta}^2)\,.
\label{dotU}
\ee
where we have used that $[\lambda_8,U_0(\vec{x\,})]=0$.
Hence $\Omega_8$ may be absorbed in a suitable
re--definition of the energy scale
$\omega\to\omega^\prime=\omega+\frac{\sqrt{3}}{2}\Omega_8$
and thus be omitted form the rotation--vibration
coupling\footnote{Some
care is required with this argument as it only applies to local
pieces in the action. For the non--local Wess--Zumino term explicit
computation also leads to the linear combination, eq.~(\ref{omega8}).}.
as in eq.~(\ref{omega8}).
However, it may not be omitted from the "classical" level, eq~(\ref{lcol})
which does not contain the fluctuation energy to compensate for $\Omega_8$.

\end{document}